\documentclass[journal]{IEEEtran}
\usepackage{cite}

\usepackage{cite}
\usepackage{amsmath,amssymb,amsfonts}
\usepackage{graphicx}
\usepackage{xcolor}
\usepackage{textcomp}
\usepackage{caption}
\usepackage{array}
\usepackage[caption=false,font=footnotesize]{subfig}

\usepackage{graphicx}
\usepackage{xcolor}
\ifCLASSINFOpdf
\else
\fi
\usepackage{amsmath}
\usepackage{amssymb}
\begin{document}
%
\title{ A Cryogenic Hybrid Photonic/CMOS Controller Architecture for Scalable Superconducting Qubit Control}
%
%
%

\author{Bowen Liu,~\IEEEmembership{Student Member,~IEEE,}
        Zhaoran Rena Huang,~\IEEEmembership{Member,~IEEE}
\thanks{Bowen Liu and Zhaoran Rena Huang are with the Department of Physics, Applied Physics, and Astronomy, Rensselaer Polytechnic Institute, Troy, NY 12180, USA}
\thanks{Zhaoran Rena Huang is with the Department of Electrical, Computer, and Systems Engineering, Rensselaer Polytechnic Institute, Troy, NY 12180, USA}
}

\maketitle

\begin{abstract}
Scaling superconducting quantum computers toward thousands of control channels remains a difficult hardware problem. Meeting this challenge requires reducing the room-temperature-to-cryogenic interconnect burden and \(4~\mathrm{K}\) power dissipation while retaining channel-specific programmability at \(4~\mathrm{K}\). This work develops a hybrid photonic/CMOS control architecture in which optical fibers distribute shared Gaussian and derivative pulse templates, while local cryogenic CMOS circuits provide event-rate amplitude programming, pulse blocking, hold-duration control, photodetection, and microwave upconversion. The architecture supports template-based XY/DRAG control and smooth-edged flat-top envelopes for two-qubit control through separate local signal paths. Compared with full-waveform Cryo-CMOS, it moves high-speed sampled-waveform generation out of each cryogenic channel. Compared with per-channel photonic waveform delivery, it retains a programmable endpoint after optical distribution for applying updated channel coefficients and control states. We evaluate the architecture using post-layout circuit simulations, optical and interconnect models, and system-level power scaling under a common \(4~\mathrm{K}\) output boundary. The modeled downlink power is below \(1~\mathrm{mW}\) per qubit-control channel at the thousand-channel scale, substantially below the normalized full-waveform Cryo-CMOS result and in the same range as the strict \(50~\Omega\) direct-detection WDM reference. Simulated \(I\)- and \(Q\)-path outputs are propagated through a five-level transmon model, while the simulated sample-and-hold output is evaluated in a coupled-transmon model. The tested single-qubit cases remain below the \(10^{-3}\) worst-state error reference, and the representative two-qubit result remains near the same scale after model-level calibration. These results support the feasibility of combining low-power photonic signal delivery with limited in-fridge programmability for template-compatible superconducting-qubit control.
\end{abstract}


\maketitle

\section{INTRODUCTION}
\label{sec:introduction}
Transmon qubits are conventionally driven by microwave pulses synthesized by room-temperature electronics and delivered to the millikelvin qubit stage through attenuated and filtered RF coaxial lines \cite{Krinner2019EPJQT,Bardin2021Microwaves}.
This room-temperature control approach has enabled superconducting processors at the hundred-qubit scale.
For example, the Zuchongzhi 3.0 processor integrates \(105\) qubits and was used to run an \(83\)-qubit, 32-cycle random-circuit-sampling experiment \cite{Gao2025PRL}.
However, in predominantly point-to-point control stacks, increasing the number of independently controlled channels generally adds coaxial lines, filters, and attenuators.
This growth creates two bottlenecks: wiring density and cold-stage heat load.
Representative dilution refrigerators provide cooling power on the order of watts at 4~K but only tens of microwatts at the mixing chamber \cite{Krinner2019EPJQT,Joshi2024JLT}.
Wiring heat and cold-stage power dissipation therefore become critical system constraints as the channel count increases.

One response is to move selected control functions into the refrigerator.
Cryo-CMOS controllers place waveform generation, sequencing, and upconversion at the 3--4~K stage, closer to the quantum processor.
Examples include the 28-nm cryogenic pulse modulator of Bardin \textit{et al.}, the 14-nm FinFET transmon controller demonstrated by IBM, and the Intel/QuTech wideband Cryo-CMOS controller for frequency-multiplexed qubit control \cite{Bardin2019JSSC,Chakraborty2022JSSC,vanDijk2020JSSC}.
Collectively, these demonstrations establish that microwave generation and a programmable control point can be placed inside the refrigerator.
The IBM semi-autonomous controller stores and executes pulse sequences at 4~K and has been used for qubit characterization, single- and two-qubit calibration, and two-qubit control \cite{Chakraborty2022JSSC,Underwood2024PRXQ}.
Complementary cryogenic readout SoCs and a 1-GS/s ADC operating at 4--4.2~K demonstrate circuit-level feasibility for moving receiver and digitization functions toward the cryogenic stage \cite{VanWinckel2022CryoReadout,Kiene2023CryoADC}.
System-level cryogenic-control proposals envision co-integrated waveform generation, readout, state discrimination, and conditional waveform generation to reduce room-temperature data transfer and shorten the readout-to-actuation path \cite{Prathapan2022WOLTE,Underwood2024PRXQ}.
These roadmaps also identify lower communication latency, reduced baseband dispersion, lower RF-delivery loss, and fewer room-temperature connections as potential system-level benefits of moving selected functions closer to the qubits \cite{Underwood2024PRXQ}.

The cost of this local functionality is active power at 4~K.
Waveform generation and sequencing require high-speed DACs, mixers, clock circuits, waveform memory, digital logic, and RF output stages.
The reported power depends strongly on which functions are included.
Bardin \textit{et al.} reported \(2~\mathrm{mW}\) of total ac and dc power for a single-channel XY pulse-modulator IC operating at 3~K under continuous \(\pi\)-pulse operation, whereas the more complete controller used by Underwood \textit{et al.} dissipated approximately \(23~\mathrm{mW/qubit}\) under active control \cite{Bardin2019JSSC,Underwood2024PRXQ}.
A superconducting-qubit readout SoC operating at 4~K reported \(1.16~\mathrm{mW/qubit}\) for the readout path alone \cite{VanWinckel2022CryoReadout}.
These results identify 4~K active-power dissipation as one principal scaling constraint of increasingly functional Cryo-CMOS integration.

Optical signal delivery provides an alternative.
Optical fibers have much lower passive thermal conduction than conventional RF coaxial cables and can carry high-bandwidth, wavelength-multiplexed signals \cite{Krinner2019EPJQT,Joshi2024JLT}.
Lecocq \textit{et al.} demonstrated superconducting-qubit control and readout using modulated optical signals delivered to a cryogenic photodetector at millikelvin temperature \cite{Lecocq2021Nature}.
Joshi and Moazeni modeled an XY-control architecture with room-temperature RF generation and optical-to-microwave conversion at 4~K, and identified wavelength-division multiplexing (WDM) as a prospective means of further reducing the physical-fiber count \cite{Joshi2024JLT}.
Under their cooling and link assumptions, the non-WDM analysis projects connectivity for thousands of XY-control channels.
Photonic links can therefore reduce passive conductive heat, while wavelength multiplexing provides a route to lower physical-fiber count.
Reduced cryogenic wiring, however, does not by itself provide a programmable control point after signal delivery.

For independently controlled transmon channels, accurate rotation angles require channel-specific amplitude calibration.
Nonlinearity in the complete control path can make the achieved rotation deviate from the value predicted by the programmed drive amplitude.
Laz{\u{a}}r \textit{et al.} experimentally characterized this nonlinear relation and corrected the resulting rotation error through amplitude calibration \cite{Lazar2023PRApplied}.
Underwood \textit{et al.} used a controller anchored at 4~K to execute qubit-calibration sequences.
Their repeated amplitude calibrations showed that the optimized DAC amplitude coefficients varied across repeated measurements \cite{Underwood2024PRXQ}.
Thus, an architecture intended to support independently calibrated rotations must provide an effective per-channel amplitude parameter in the control stack and a mechanism to refresh it when recalibration is required.

Both full-waveform Cryo-CMOS and per-channel WDM control can support channel-specific amplitude calibration.
The IBM controller demonstrates the Cryo-CMOS case: waveform and instruction memories reside at 4~K, and calibrated amplitude parameters are applied during local waveform generation \cite{Chakraborty2022JSSC,Underwood2024PRXQ}.
In the WDM reference modeled here, the calibrated coefficient is applied at room temperature to the corresponding waveform or modulator drive before optical transmission \cite{Joshi2024JLT}.
This implementation supports independent amplitude calibration, but the updated coefficient is applied upstream of the cryogenic receiver.

Latency becomes a concern when an update is conditioned on a recent qubit measurement.
Periodic recalibration compensates changes that occur over successive experiments and only requires the new coefficient to be available before it is next used.
Measurement-conditioned control instead requires a short readout-to-actuation path.
Salath{\'e} \textit{et al.} reported an analog-input-to-feedback-trigger latency of \(110\pm3~\mathrm{ns}\) in a room-temperature FPGA system and inferred an additional \(69\pm7~\mathrm{ns}\) group delay from cables and analog components \cite{Salathe2018PRApplied}.
When readout processing, decision logic, and waveform modulation remain at room temperature, the conditional command and the resulting control signal must traverse the room-temperature decision and modulation path before returning to the refrigerator.
A cryogenic controller can instead provide a local actuation endpoint when combined with suitable state-discrimination and decision logic \cite{Prathapan2022WOLTE}.

Motivated by these requirements, this work asks whether channel-specific programmability can be introduced at 4~K while retaining the low cold-stage dissipation sought by photonic signal delivery, and what additional power and interconnect costs are incurred in doing so.
To answer this question, we propose and evaluate a hybrid photonic/CMOS controller for template-compatible channel groups, in which each group shares a small set of high-bandwidth pulse templates while each channel retains a small set of low-rate local control states.
Gaussian envelopes and derivative quadratures are established transmon-control pulse components and serve as the two shared optical templates considered here \cite{Motzoi2009PRL,Hyyppa2024PRXQ}.
The templates are generated at room temperature, delivered to 4~K, and distributed through passive optical fan-out.
On each output branch, a low-speed DAC drives an active microring modulator (MRM)\footnote{In this paper, MRM denotes the active microring modulator used in the proposed controller, whereas MRR denotes the passive microring resonator used for wavelength selection in the WDM reference.} to set the calibrated channel amplitude or suppress a selected pulse.
A photodetector recovers the electrical envelope, and shared local oscillators(LO) support I/Q upconversion.
For two-qubit-control applications, the I/Q path can carry template-based microwave envelopes, while the sample-and-hold(S/H) path can form smooth-edged flat-top baseband envelopes.
Across these paths, amplitude and pulse-enable states remain channel specific, and the flat-top duration is set locally.
This partition avoids the need for a complete high-speed sampled-waveform generator in every cryogenic channel.

For a consistent power comparison, the proposed controller, full-waveform Cryo-CMOS, and per-channel WDM photonic control are evaluated at the same \(4~\mathrm{K}\) electrical output boundary. The system analysis accounts for architecture-specific optical and RF losses, local optical fan-out, template-fiber and LO-feed scaling, and the resulting \(4~\mathrm{K}\) power breakdown. Separately simulated \(I\)- and \(Q\)-path waveforms are combined in a five-level transmon model to assess representative XY/DRAG control-chain errors against a \(10^{-3}\) screening target. The simulated S/H output waveform is evaluated separately in a coupled-transmon model to quantify the effect of envelope distortion on a representative flux-controlled CZ operation.
Together, these analyses characterize the proposed architecture at the power, interconnect, and control-waveform levels.
We also identify the remaining implementation requirements and experimental steps needed for device-specific calibration and validation.

Table~\ref{tab:sota_comparison} compares representative cryogenic electronic and photonic control approaches in terms of waveform generation, local programmability, cold-stage power, interconnect scaling, storage, and timing. Values from prior work retain their published operating conditions and reference planes. 

\begin{table*}[!t]
\centering
\scriptsize
\setlength{\tabcolsep}{1.5pt}
\renewcommand{\arraystretch}{1.10}
\caption{Representative cryogenic electronic and photonic qubit-control approaches. Source-specific validation conditions, workloads, temperatures, and reference planes are retained; NR denotes not reported.}
\label{tab:sota_comparison}
\begin{tabular}{@{}
>{\raggedright\arraybackslash}p{0.145\textwidth}
>{\raggedright\arraybackslash}p{0.130\textwidth}
>{\raggedright\arraybackslash}p{0.120\textwidth}
>{\raggedright\arraybackslash}p{0.150\textwidth}
>{\raggedright\arraybackslash}p{0.115\textwidth}
>{\raggedright\arraybackslash}p{0.135\textwidth}
>{\raggedright\arraybackslash}p{0.125\textwidth}@{}}
\hline
\textbf{Architecture and validation} &
\textbf{Waveform generation} &
\textbf{Cold-stage control} &
\textbf{Cold-stage power / operating point} &
\textbf{Interconnect or scale} &
\textbf{Cold-stage storage} &
\textbf{Timing evidence} \\
\hline

\textbf{28-nm Cryo-CMOS pulse modulator}
\cite{Bardin2019JSSC}\newline
Fabricated at 3~K; XY control &
Local 4--8-GHz XY-pulse generation using selectable symmetric envelopes &
Envelope selection, amplitude, polarity, and carrier phase &
Less than \(2~\mathrm{mW}\) per IC during continuous \(\pi\)-pulse operation &
External I/Q LO, clock, SPI, trigger, and supplies; one RF output &
16 envelope configurations; total storage capacity NR &
Pulse-controller clock up to \(2.1~\mathrm{GHz}\); trigger-to-output latency NR \\

\textbf{22-nm frequency-multiplexed Cryo-CMOS}
\cite{vanDijk2020JSSC}\newline
Fabricated at 3~K; RF tests and spin-qubit control &
Local 2--20-GHz DDS/SSB generation; up to 32 frequency-multiplexed qubits per RF output &
Frequency, phase, amplitude, and instruction selection &
\(12~\mathrm{mW/qubit}\) at a \(1~\mathrm{GHz}\) clock under the reported 32-qubits-per-output normalization &
Shared SPI and trigger; external LO, clock, and supplies &
4096 instructions and more than 40000 waveform points &
Approximately \(75~\mathrm{ns}\) minimum interval between external instruction triggers; end-to-end latency NR \\

\textbf{28-nm full unit-cell Cryo-CMOS}
\cite{Yoo2023JSSC}\newline
Fabricated at 3~K; two-transmon experiments &
Two 4--8-GHz XY generators and three baseband qubit/coupler-control generators &
Waveform selection, XY enable, duration, amplitude, polarity, and phase &
\(7.4~\mathrm{mW}\) for the complete two-qubit-control IC at high activity; approximately \(4~\mathrm{mW/qubit}\) in the demonstrated configuration &
External LO, \(1~\mathrm{GHz}\) clock, SPI, trigger, and supplies; physical feed count NR &
One 16-entry waveform memory per signal generator; 26-bit instruction; up to 512 sequence steps &
\(1~\mathrm{GHz}\) system clock and \(2~\mathrm{ns}\) minimum inter-pulse dead time; end-to-end latency NR \\

\textbf{14-nm semi-autonomous Cryo-CMOS}
\cite{Chakraborty2022JSSC,Underwood2024PRXQ}\newline
Fabricated at 4~K; single- and two-transmon experiments &
Stored I/Q waveforms and sequences for single-qubit and CR control &
Amplitude, frequency, phase/frame, duration, and sequence &
Approximately \(23~\mathrm{mW/qubit}\) measured under active control &
External programming, clock, LO, references, and supplies; physical feed count NR &
32-kB instruction, 20-kB waveform, and 32-kB data memories &
Triggered execution of preloaded sequences; end-to-end latency NR \\

\textbf{Direct cryogenic photonic link}
\cite{Lecocq2021Nature}\newline
Photodetector at 20~mK; transmon control and readout &
Complete room-temperature microwave waveform transported optically &
No programmable control state after photodetection &
Total link power NR; approximately \(4~\mu\mathrm{A}\) average detector current for representative control &
One fiber/photodetector link demonstrated; system-scale optical multiplexing not demonstrated &
No cold-stage waveform memory in the demonstrated link &
NR \\

\textbf{4-K RF-photonic control model}
\cite{Joshi2024JLT}\newline
Analytical XY-link and thermal-scaling study &
Complete room-temperature XY waveform transported optically &
No programmable control state after photodetection &
Ideal modeled \(1.4~\mu\mathrm{A}\) detector operating point and \(5.6~\mu\mathrm{W/fiber}\) passive RT-to-4-K heat &
Approximately 3000 non-WDM XY lines projected under the adopted assumptions; WDM proposed &
No cold-stage waveform memory in the model &
NR \\

\textbf{Proposed shared-template photonic/CMOS}\newline
Circuit-simulated electrical blocks, analytical system scaling, and five-level and coupled-transmon analyses &
Shared Gaussian/derivative templates and a locally formed smooth-edged hold envelope &
8-bit I/Q/Z amplitudes, pulse enable/block, and hold duration &
Modeled \(0.905~\mathrm{mW/channel}\) at \(N=1000\) and the defined 4-K boundary; local digital-state and timing power excluded &
126 template fibers plus four LO coaxes at \(N=1000\), \(G=16\); auxiliary feeds excluded &
24-bit resident amplitude state, plus pulse-enable and implementation-dependent hold/timing fields; no sampled-waveform store &
\(100~\mathrm{MHz}\) coefficient-DAC update rate; complete trigger-to-output latency not extracted \\

\hline
\end{tabular}
\end{table*}

\section{Control Requirements and Template Scope}
\label{sec:background}

\subsection{Control Frequencies, Timing, and Reference Plane}

Superconducting transmon gates are implemented using calibrated microwave or baseband pulses applied to the corresponding control lines.
For XY control, a microwave tone near the qubit transition frequency drives the \(\lvert 0\rangle\leftrightarrow\lvert 1\rangle\) transition.
The carrier frequency selects the addressed transition, the carrier phase sets the rotation axis in the equatorial plane, and the envelope amplitude and duration determine the pulse area and hence the rotation angle \cite{Bardin2021Microwaves,Joshi2024JLT}.
Representative transmon frequencies lie in the \(4\)--\(8~\mathrm{GHz}\) range, and single-qubit pulses commonly have durations of approximately \(10\)--\(30~\mathrm{ns}\) \cite{Joshi2024JLT}.

In a multiqubit processor, transition frequencies are assigned according to an engineered frequency plan.
Neighboring qubits are generally detuned to reduce frequency collisions and crosstalk, while frequency classes can be reused across sufficiently separated qubits.
Representative surface-code layouts use three or four repeated single-qubit frequency classes rather than one unique carrier per physical qubit \cite{Asaad2016Broadcasting,Versluis2017SurfaceCode}.
The four-class LO network used in Sec.~\ref{sec:proposed_architecture} is therefore a representative frequency-reuse configuration.

Two-qubit control introduces longer and gate-dependent time scales.
Microwave cross-resonance or echoed cross-resonance (CR/ECR) pulses are commonly approximately \(100\)--\(500~\mathrm{ns}\), whereas reported flux-controlled CZ pulses range from tens of nanoseconds to approximately \(100~\mathrm{ns}\) \cite{Sheldon2016PRA,Underwood2024PRXQ,Li2024PRX}.
These ranges motivate a locally programmable hold interval for the smooth-edged flat-top envelope generated by the S/H path.

The voltage required for a calibrated gate is not a universal transmon parameter.
It depends on the control-line coupling, downstream attenuation, package loss, on-chip routing, pulse shape, and gate duration.

\subsection{Pulse Families and Template Scope}

\begin{figure}[t]
    \centering

    \begin{minipage}{0.48\columnwidth}
        \centering
        \fbox{
        \includegraphics[
            width=\linewidth,
            trim={1cm 1.3cm 1cm 1.3cm},
            clip
        ]{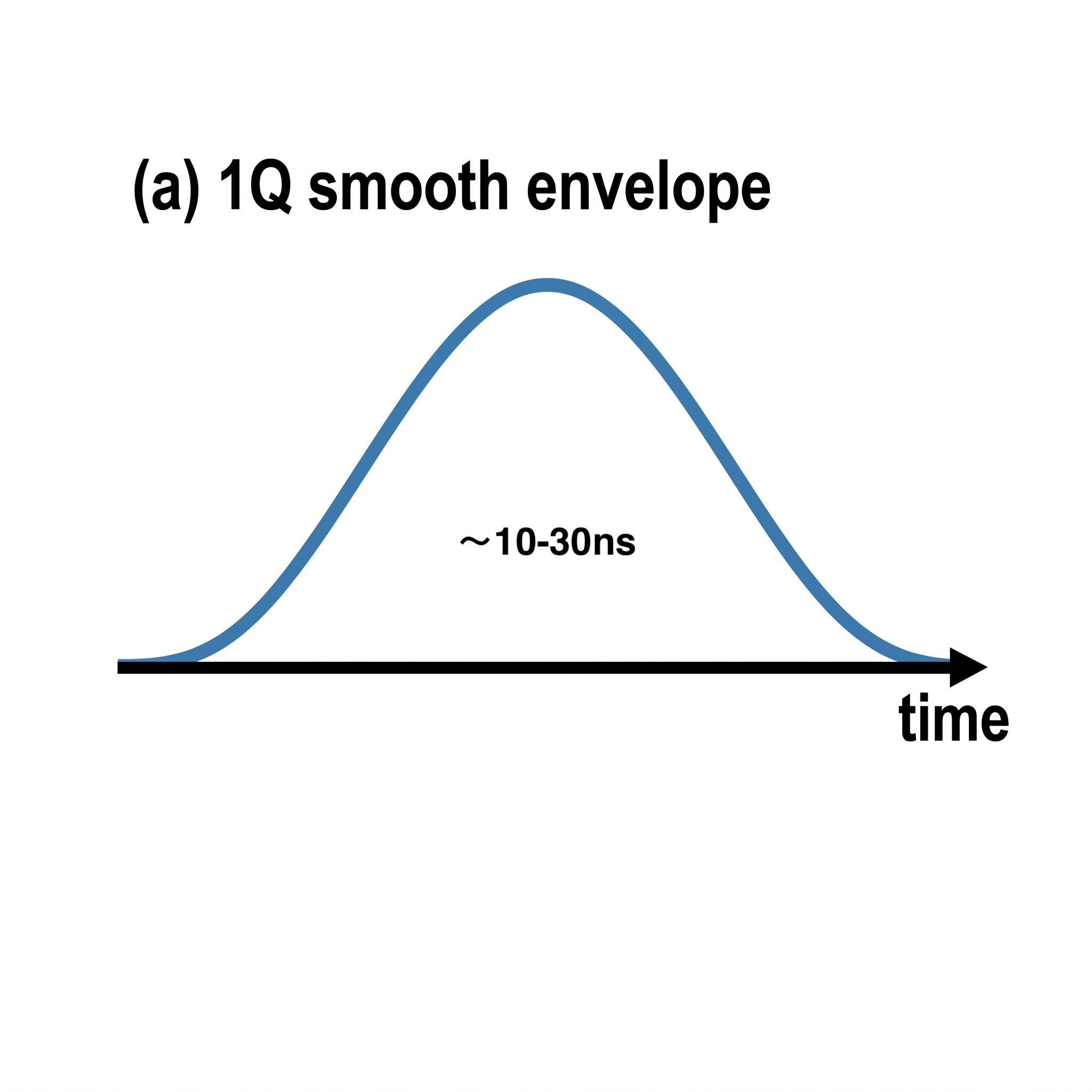}}
    \end{minipage}
    \hfill
    \begin{minipage}{0.48\columnwidth}
        \centering
        \fbox{
        \includegraphics[
            width=\linewidth,
            trim={1cm 1.3cm 1cm 1.3cm},
            clip
        ]{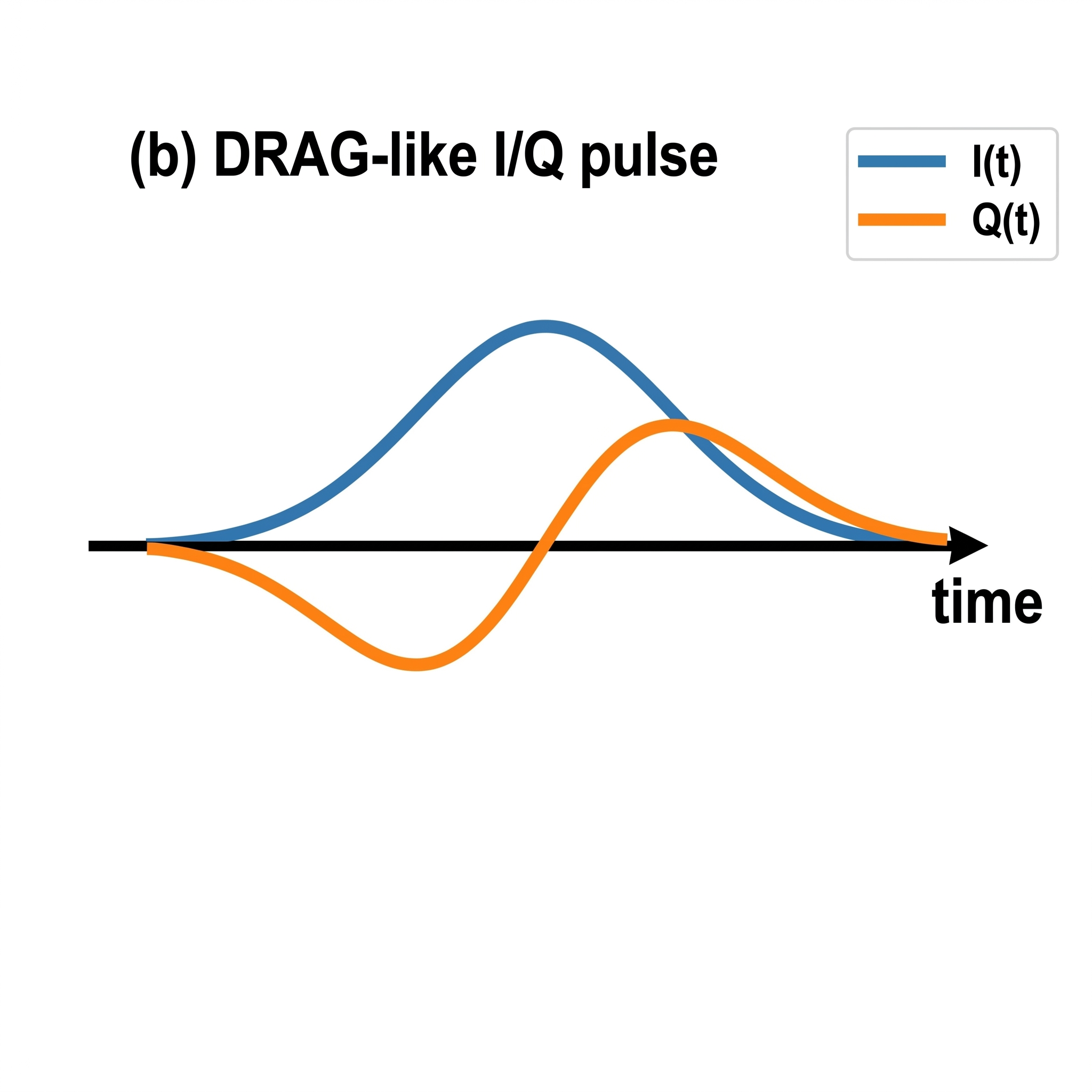}}
    \end{minipage}

    \vspace{0.35em}

    \begin{minipage}{0.48\columnwidth}
        \centering
        \fbox{
        \includegraphics[
            width=\linewidth,
            trim={1cm 1.3cm 1cm 1.3cm},
            clip
        ]{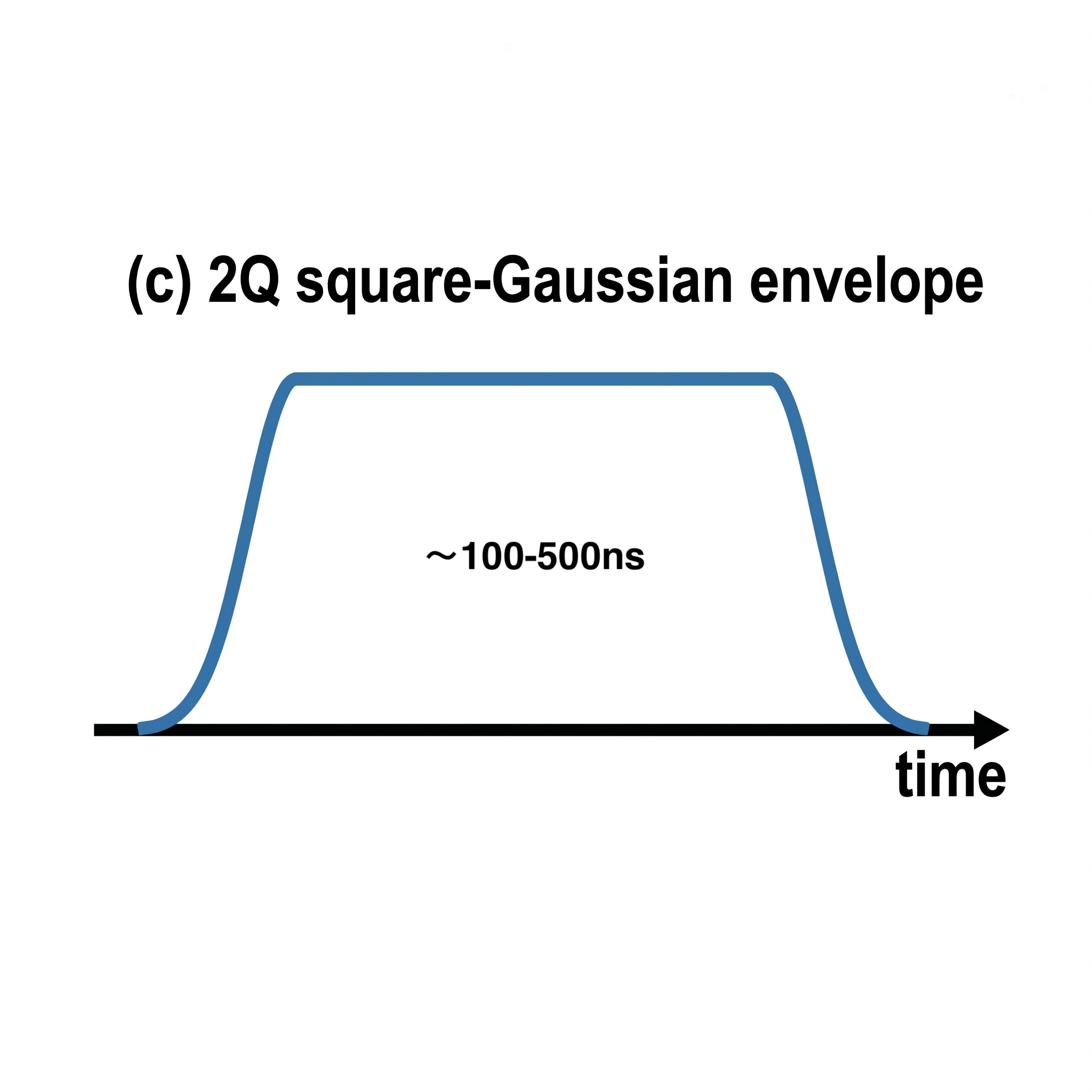}}
    \end{minipage}
    \hfill
    \begin{minipage}{0.48\columnwidth}
        \centering
        \fbox{
        \includegraphics[
            width=\linewidth,
            trim={1cm 1.3cm 1cm 1.3cm},
            clip
        ]{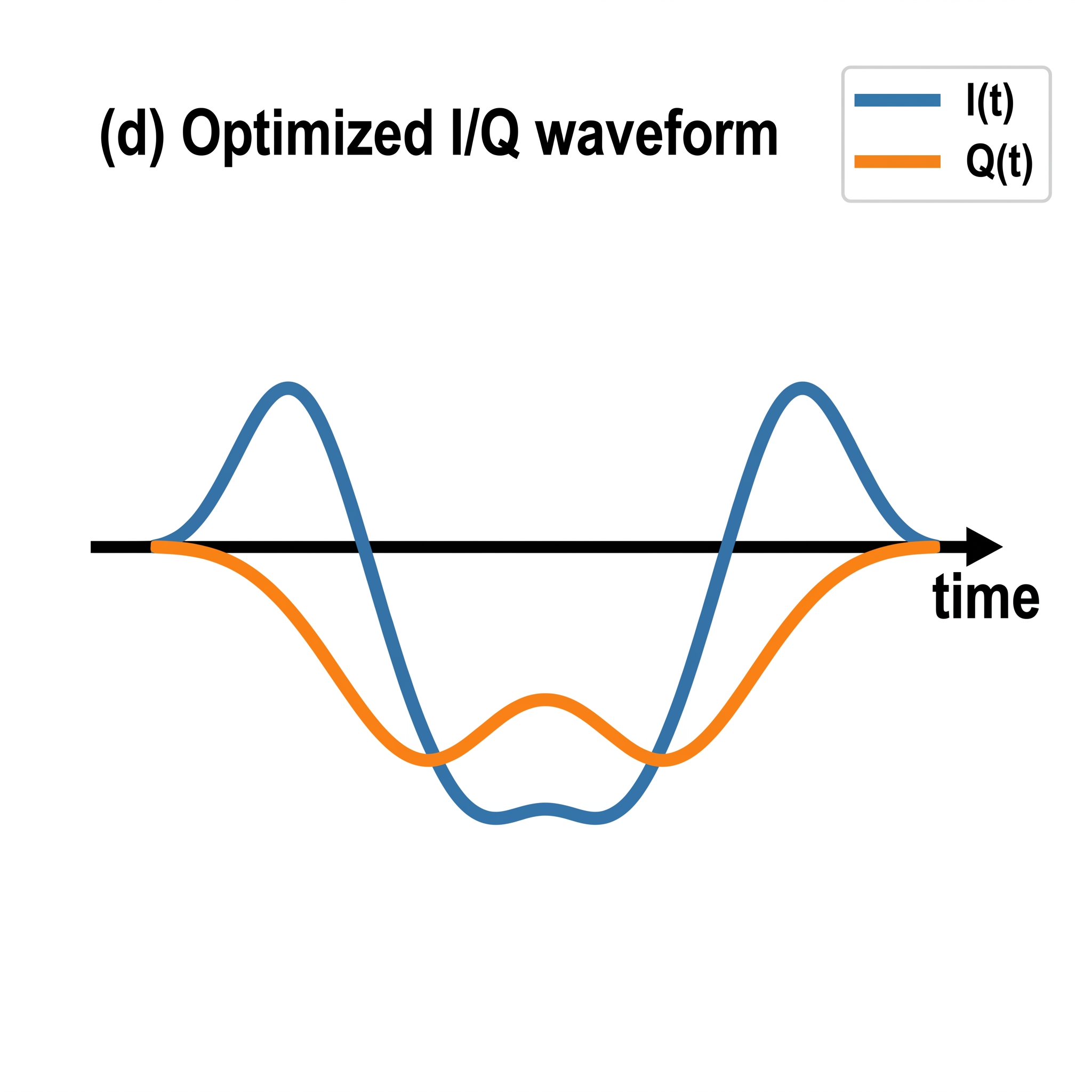}}
    \end{minipage}

    \caption{Schematic superconducting-qubit control envelopes: (a) a smooth single-qubit envelope; (b) a DRAG-like I/Q pulse; (c) a smooth-edged flat-top envelope representative of two-qubit-control pulse families; and (d) an optimized sample-by-sample I/Q waveform shown as an out-of-scope full-waveform reference. For panel (c), microwave CR/ECR and baseband flux-control implementations can use qualitatively similar envelopes, but their durations, amplitudes, downstream signal paths, and gate calibrations are different.}
    \label{fig:pulse_shapes}
\end{figure}

Fig.~\ref{fig:pulse_shapes} summarizes four representative waveform classes.
Panels (a)--(c) motivate the template-compatible paths considered in this work, whereas panel (d) illustrates sample-by-sample waveform flexibility that is not retained by the quantitative two-template model.

For single-qubit XY gates, Gaussian or raised-cosine envelopes reduce spectral content outside the intended control band compared with square pulses.
Derivative-removal-by-adiabatic-gate (DRAG) control adds a quadrature component proportional to the time derivative of the in-phase envelope to suppress unwanted excitation of the transmon's $(\lvert 2\rangle)$ and higher energy levels \cite{Motzoi2009PRL,Hyyppa2024PRXQ}.
In a calibrated DRAG implementation, this relation can be written as \(Q(t)=\beta\dot{I}(t)\), where the magnitude and sign of \(\beta\) depend on the device and waveform convention.
The Gaussian and derivative envelopes form the two shared template families used in this quantitative model.
For a fixed-duration gate, the calibrated envelope shape can remain unchanged while the amplitude and pulse-enable state remain channel specific.
Laz{\u{a}}r \textit{et al.} demonstrated arbitrary-angle gates by calibrating the nonlinear relation between programmed amplitude and rotation angle for a fixed \(15~\mathrm{ns}\) pulse \cite{Lazar2023PRApplied}.

Sample-by-sample optimized I/Q pulses constitute a separate waveform class.
Optimal-control methods can tailor these pulses to a particular device and gate by incorporating bandwidth, noise, and crosstalk constraints \cite{Carvalho2021PRApplied}.
Fig.~\ref{fig:pulse_shapes}(d) illustrates this reference case. Supporting such waveforms in the proposed architecture would require a larger template basis or complete waveform synthesis. 

Smooth-edged flat-top square-Gaussian envelopes in Fig.~\ref{fig:pulse_shapes}(c) are commonly used for two-qubit control.
In microwave-driven CR/ECR gates, this envelope modulates a carrier near the target-qubit transition frequency; echoed segments and cancellation tones may also be applied to suppress unwanted interaction terms \cite{Sheldon2016PRA,Heya2021PRXQ,Underwood2024PRXQ}.
In flux-controlled CZ gates, a baseband square-Gaussian pulse tunes a qubit or coupler into the desired interaction region; the flat-top interval accumulates the conditional phase, while the smooth edges reduce nonadiabatic leakage during the frequency excursion \cite{Li2024PRX,Glaser2025PRApplied}.

\subsection{Control Accuracy and Fidelity Mapping}

The controller must preserve the amplitude, phase, timing, and frequency accuracy required by calibrated gates so that it does not become the dominant source of gate error.
As a reference point, early superconducting-qubit surface-code-threshold experiments reported an average single-qubit gate fidelity of \(99.92\%\) and two-qubit gate fidelity up to \(99.4\%\) \cite{Barends2014Nature}.
More recent demonstrations have reported single-qubit fidelities near \(99.98\%\) and two-qubit gate fidelities of approximately \(99.9\%\) in optimized devices \cite{Li2024PRX}.
These measured gate errors include contributions from the quantum device, decoherence, leakage, calibration, and the classical control chain.
The controller-induced contribution must therefore remain below the total physical-operation error budget.

To assess the control-chain contribution, we follow the carrier/envelope decomposition of van Dijk \textit{et al.} \cite{vanDijk2019PRApplied}.
Carrier-frequency error, carrier-phase error, phase noise, RF feedthrough, and crosstalk affect the rotation axis or introduce unwanted drive.
Envelope-gain error, pulse-width error, template delay, detector uncertainty, and incomplete pulse blocking alter the pulse area or introduce an unwanted envelope.
Following the \(99.9\%\) operation-fidelity target used by van Dijk \textit{et al.}, the subsequent analysis uses a controller-induced infidelity of \(10^{-3}\) as a common screening target for representative carrier- and envelope-path nonidealities.

\section{Shared-Template Photonic/CMOS Control Architecture}
\label{sec:proposed_architecture}
\subsection{System Partition and Signal Paths}
\label{subsec:architecture_overview}

\begin{figure}[!t]
    \centering
    \includegraphics[width=\columnwidth]{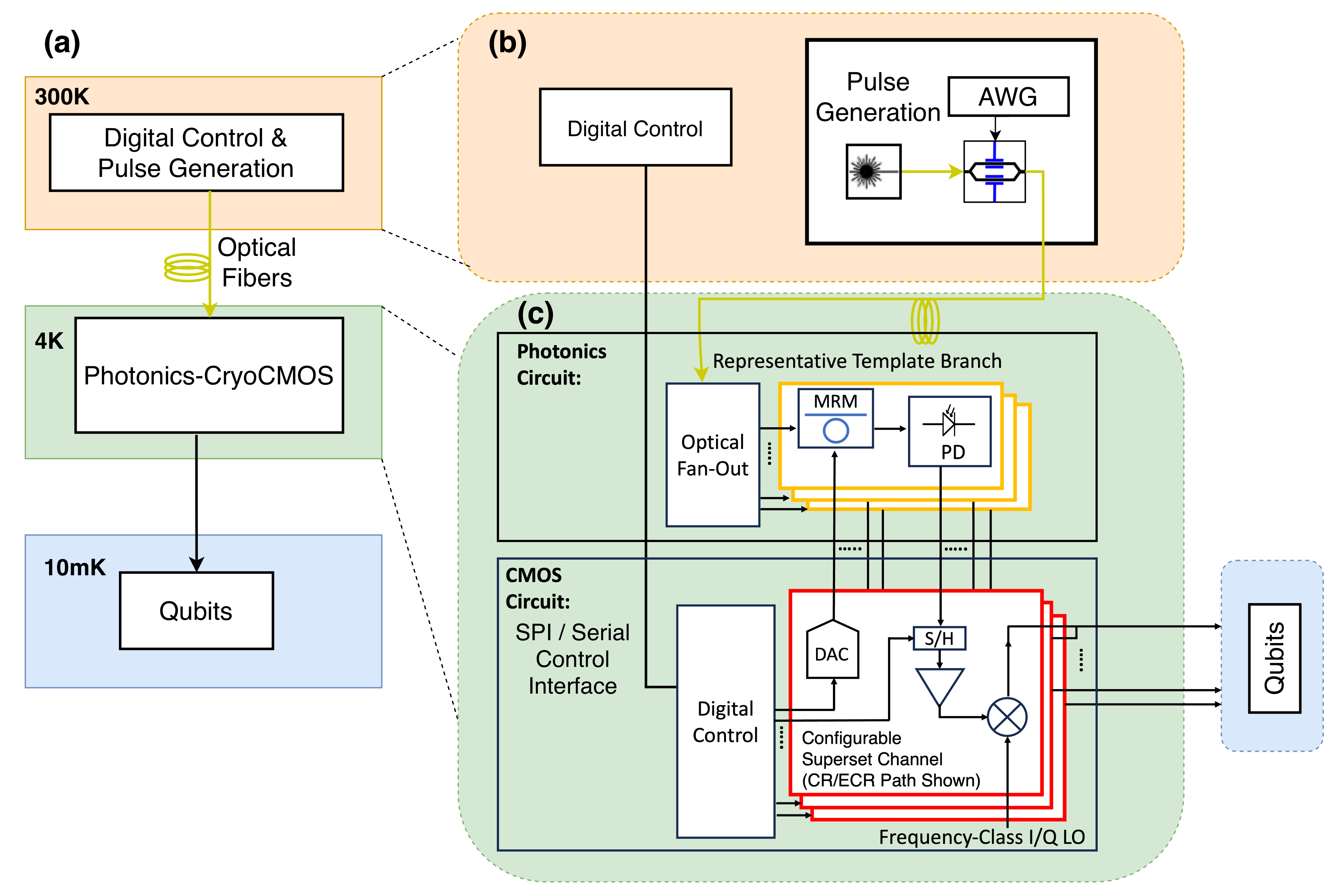}
    \caption{Proposed hybrid cryogenic photonic/CMOS control architecture. (a) System-level partitioning across room temperature, the 4~K controller stage, and the millikelvin qubit stage. (b) Room-temperature optical-template generation and low-rate digital configuration. (c) A representative configurable 4~K photonic/CMOS channel containing the local optical-amplitude-control, photodetection, S/H, and microwave-upconversion blocks. The illustrated signal chain is a functional superset that includes an optional microwave CR/ECR envelope path. It is not a mandatory serial path for every gate. Single-qubit XY/DRAG control bypasses the S/H stage, whereas baseband Z/CZ control bypasses the LO mixer. One representative optical branch is shown; the matched quadrature branch, the signed-\(Q\) polarity switch, and gate-dependent bypass connections are omitted for clarity. }
    \label{fig:proposed_architecture}
\end{figure}

A schematic of the proposed hybrid photonic/CMOS controller is shown in Fig.~\ref{fig:proposed_architecture}.
The architecture separates shared high-bandwidth template generation at room temperature from channel-specific amplitude, pulse-selection, and timing control at 4~K.
The optical templates are delivered to local channel groups at 4~K, where the photonic/CMOS circuitry applies the programmed amplitude, pulse-enable, and timing states and converts the selected templates into microwave or baseband control signals for the millikelvin qubit stage.

The programming interface originates at room temperature, while the programmable state and execution logic reside in the 4~K controller.
A low-rate serial peripheral interface (SPI) loads per-channel amplitude codes, pulse-enable masks, S/H duration parameters, and event-state or sequence entries into local registers.
During circuit execution, the 4~K control logic selects and applies these stored states at the scheduled optical-template slots.
The SPI link therefore carries low-rate configuration words rather than high-rate waveform samples.

The nominal implementation uses two shared optical-template rails per local group: one Gaussian rail that is routed to the main I-envelope branch and the modeled baseband Z/CZ branch, and one derivative/DRAG rail for the Q-envelope branch. At room temperature, a shared optical-source/modulator subsystem generates the Gaussian and derivative template trains under a common scheduling reference. After entering the 4~K stage, the templates are redistributed through local passive optical networks.

Larger systems replicate these local optical groups. Each endpoint branch contains an electrically driven MRM controlled by local CMOS circuitry. Intermediate transmission states set the calibrated pulse amplitude, and the pass and block states select scheduled template slots. The MRM is updated at gate-event boundaries, not at the sample rate of the optical template. This avoids a per-channel cryogenic radio-frequency arbitrary-waveform generator (RF-AWG) while retaining pulse selection and amplitude control.

After local optical programming, a photodetector (PD) converts each selected optical template into an electrical envelope at the 4~K stage. The downstream signal path is selected according to the gate type. For single-qubit XY/DRAG control, the Gaussian and derivative envelopes drive the I and Q microwave mixer paths without using the S/H stage. For an optional microwave-driven CR/ECR extension, the S/H stage can extend a detected Gaussian edge to form a square-Gaussian-like envelope before microwave upconversion. For baseband Z/CZ control, the S/H and output-buffer path forms the modeled flux-control envelope without passing through the LO mixer. 

For logical microwave-control channel \(k\), let \(c(k)\) denote its assigned LO frequency class. The ideal XY/DRAG path is represented by
\begin{equation}
\begin{aligned}
v_k^{\mathrm{RF}}(t)={}&I_k(t)\cos\!\left(2\pi f_{\mathrm{LO},c(k)}t\right)\\
&+Q_k(t)\sin\!\left(2\pi f_{\mathrm{LO},c(k)}t\right),\\
Q_k(t)={}&\beta_k\frac{dI_k(t)}{dt},
\end{aligned}
\label{eq:iq_drag_path}
\end{equation}
where \(I_k(t)\) and \(Q_k(t)\) are the local baseband envelopes, \(f_{\mathrm{LO},c(k)}\) is the assigned carrier, and \(\beta_k\) is the calibrated DRAG coefficient \cite{Motzoi2009PRL,Herrmann2022Upconversion,Underwood2024PRXQ}.

Single-ended direct detection of an intensity-modulated derivative rail produces a nonnegative photocurrent. The rail therefore carries the envelope magnitude, and a pulse-synchronous signed-\(Q\) polarity switch (commutator), driven by the known template sign, restores the bipolar \(Q_k(t)\) envelope before the Q mixer. The present Fig.~\ref{fig:proposed_architecture} does not include this switch, but the switch overhead is included in the power model in later section.

\begin{figure}[!t]
    \centering
    \includegraphics[width=0.88\columnwidth]{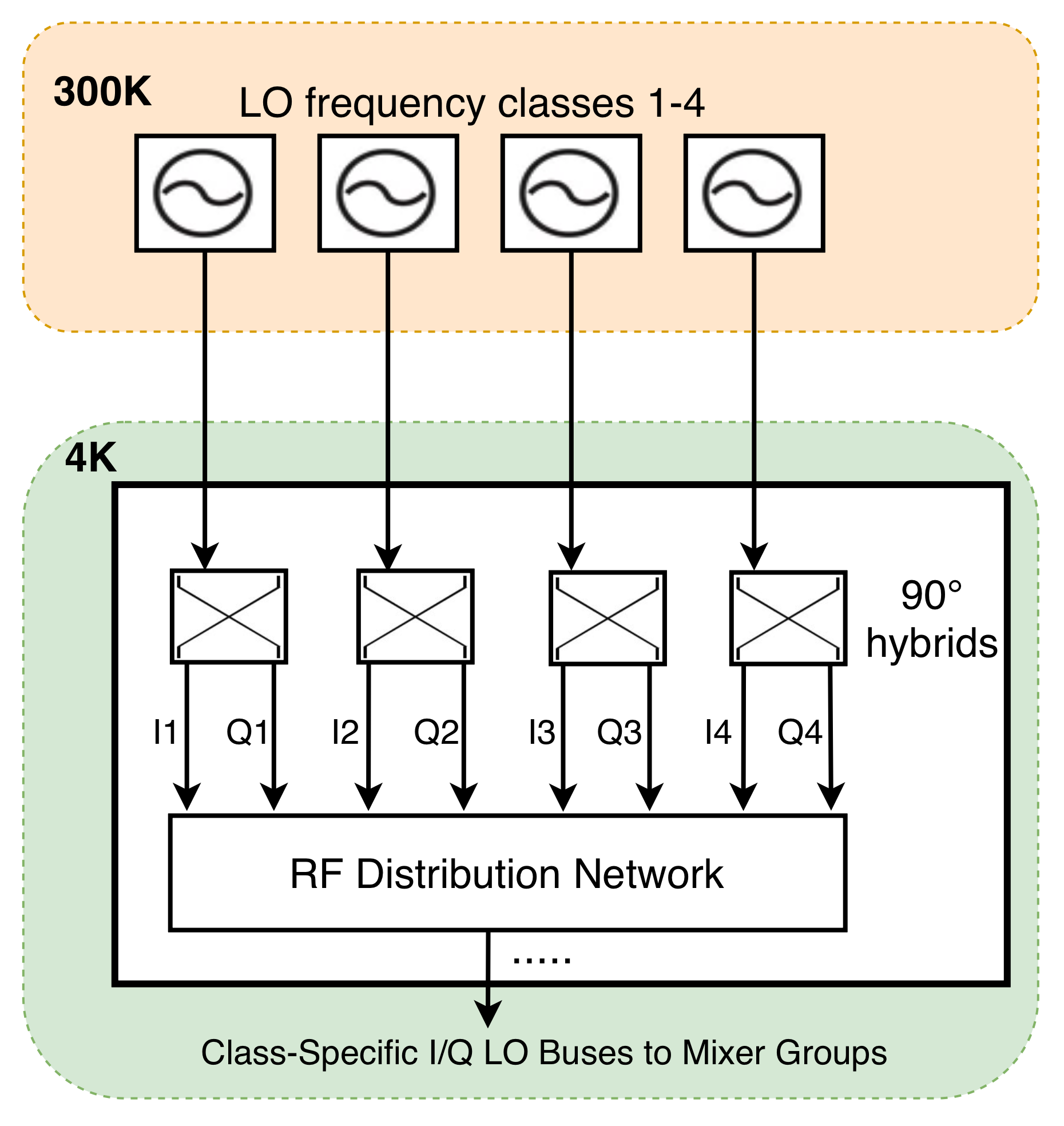}
    \caption{Shared frequency-class LO generation and distribution for the modeled XY/DRAG network. Four LO carriers phase-locked to a common reference are generated at room temperature and delivered to the 4~K stage through four shared coaxial lines. Each carrier drives one shared \(90^\circ\) quadrature hybrid. The two outputs form the in-phase and quadrature buses for that frequency class. The RF distribution block represents eight distinct class-specific buses. }
    \label{fig:lo_distribution}
\end{figure}

Fig.~\ref{fig:lo_distribution} shows the shared frequency-class LO generation and distribution network used by the baseline XY/DRAG path. The LO network supports single-qubit XY/DRAG control using four shared frequency classes, denoted \(f_{\mathrm{LO},1}\) through \(f_{\mathrm{LO},4}\). Four classes are selected as a representative frequency-reuse configuration based on tiled layouts that reuse three or four carrier classes across a processor \cite{Asaad2016Broadcasting,Versluis2017SurfaceCode}; the number is not universal.

The four room-temperature carriers are phase-locked to the same master reference used by the optical-template scheduler and are delivered to 4~K through four shared coaxial lines. This configuration avoids a cryogenic tone-selection network. At 4~K, each frequency-class carrier first drives one shared \(90^\circ\) quadrature hybrid. The two outputs of each hybrid form the in-phase and quadrature LO buses for that frequency class. Together, the four hybrids generate eight class-specific LO buses. The RF distribution network routes these buses to the mixer groups assigned to the corresponding frequency classes.

Qubits that share the same LO frequency remain independently addressable because their optical envelopes, amplitudes, scheduled-slot selection, and pulse-blocking states are programmed separately after optical distribution. In the baseline configuration, the Gaussian and derivative templates implement a calibrated fixed-axis DRAG primitive. Other logical equatorial rotation axes can be synthesized through virtual-\(Z\) frame tracking and gate decomposition. Direct per-channel selection among arbitrary physical carrier phases would require programmable I/Q routing beyond the fixed pulse-synchronous polarity switch assumed above. The proposed configuration therefore does not require a per-channel numerically controlled oscillator (NCO), a per-channel LO-tone selector, or an arbitrary-phase DAC.

\subsection{Template-Based Gate Operation and Event-Level State}
\label{subsec:template-gate-operation}

\begin{figure}[!t]
    \centering
    \includegraphics[width=\columnwidth]{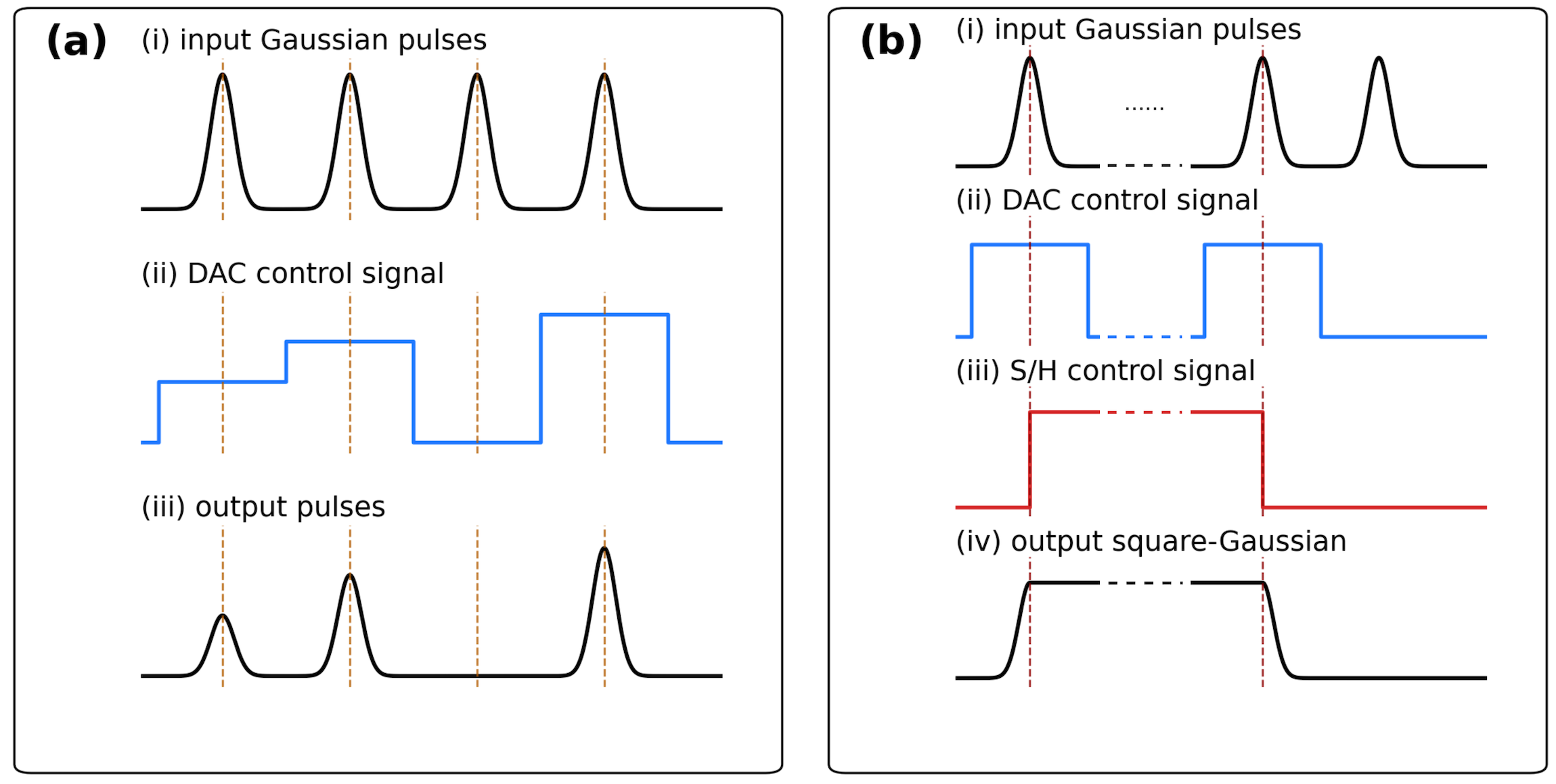}
    \caption{Local control of shared optical-template trains. (a) Single-qubit operation for one representative Gaussian branch: a preloaded local DAC code sets the channel amplitude, while pulse-enable logic passes scheduled template slots and blocks unscheduled slots. The matched derivative/Q branch and its polarity switch are omitted for clarity. (b) Formation of a smooth-edged flat-top two-qubit envelope: the first Gaussian edge event charges the envelope node, the S/H window maintains the programmed level for a gate-dependent hold interval \(T_{\mathrm{hold}}\), and a controlled release/reset event defines the falling transition. The resulting envelope can be routed to microwave upconversion for a CR/ECR extension or used directly by the baseband Z/CZ path.}
    \label{fig:local-pulse-control}
\end{figure}

Fig.~\ref{fig:local-pulse-control} illustrates how a shared optical-template train is converted into channel-specific gate envelopes.

For single-qubit control, a preloaded local DAC code sets the calibrated transmission level of the channel MRM.
Local pulse-enable logic places the MRM in the programmed transmission state for a scheduled template slot and in a near-zero-transmission state for an unscheduled slot.
The selected Gaussian pulse is therefore amplitude-scaled or blocked without reproducing its high-bandwidth waveform samples at 4~K.
For DRAG control, the same operation is applied to the matched derivative-template branch, which provides the calibrated Q-envelope magnitude before polarity restoration.

For the smooth-edged flat-top envelope shown in Fig.~\ref{fig:local-pulse-control}(b), the first detected Gaussian edge establishes the turn-on transition and charges the hold node to the programmed amplitude.
The S/H circuit maintains this level for the calibrated interaction interval.
At the end of the gate window, a controlled release or reset returns the envelope toward zero and defines the falling transition.
The hold and release circuitry can introduce hold-state droop and a charge-injection transient near the falling edge; their effects are evaluated later using the simulated S/H output waveform.

The S/H output can be routed to microwave upconversion for an optional CR/ECR extension \cite{Underwood2024PRXQ} or directly to the baseband Z/CZ output used in the quantitative two-qubit analysis \cite{Li2024PRX}.
For the flat-top envelope primitive considered here, the optical template supplies the high-bandwidth edge information, while the 4~K circuit stores only the channel-specific amplitude, pulse-enable state, and hold duration.

The template-based operation also changes the local waveform representation.
A directly stored \(M\)-sample envelope with \(b\) bits per sample requires \(Mb\) bits for a scalar path and \(2Mb\) bits for an I/Q path.
The proposed controller instead stores a template identifier together with event-level amplitude, pulse-enable, and hold or timing states, while the high-rate waveform samples are supplied optically.
Actual memory savings depend on the instruction format and waveform-reuse strategy; the present comparison therefore addresses representation scaling only.

\subsection{Device-Level Implementation Considerations}
\label{subsec:device-considerations}

The reference implementation uses a monolithic silicon-photonic/CMOS platform to keep the electrical connections among the modulators, photodetectors, and local CMOS circuits short.
GlobalFoundries 45SPCLO provides a representative 45-nm SOI platform with CMOS and silicon-photonic device libraries, including waveguides, modulators, Ge photodetectors, and fiber-coupling structures \cite{GF45SPCLO_CMC}.
A heterogeneous multi-die implementation remains an alternative if a single process cannot provide the required modulation efficiency, detector responsivity, and RF performance, although it introduces additional inter-die parasitics and packaging complexity.

The local optical-amplitude actuator is a resonant MRM, selected for its compact footprint and small electrical load.
A nonresonant Mach--Zehnder modulator would reduce sensitivity to resonance alignment but would require a longer phase-shifting path and a larger branch footprint.
At cryogenic temperature, carrier freeze-out and the required doping level affect junction resistance, optical absorption, and resonator quality factor \cite{Gehl2017Optica}.
Gevorgyan \textit{et al.} demonstrated an O-band vertical-junction spoked-ring modulator at approximately \(4.5~\mathrm{K}\), with a resonance-shift efficiency of \(46~\mathrm{GHz/V}\), a \(140~\mathrm{GHz}\) nonthermal tuning range, and on--off-keying operation up to \(10~\mathrm{Gb/s}\) using a drive of \(\leq 1~\mathrm{V_{pp}}\) \cite{gevorgyan2021cryoRing}.
This result provides a direct cryogenic precedent for sub-\(1~\mathrm{V_{pp}}\) resonant modulation and local optical-amplitude control.
Because conventional thermo-optic tuning becomes inefficient at low temperature, cooldown shifts and static device variation require calibrated electrical biasing, nonvolatile trimming, or another low-static-power tuning method \cite{adya2025nvRing}.

The reference photodetector is a waveguide-integrated Ge PIN device compatible with the monolithic platform.
Measurements of foundry-compatible Ge photodetectors at \(4.2~\mathrm{K}\) have reported O-band responsivities of approximately \(0.04\)--\(0.15~\mathrm{A/W}\), depending on detector geometry and the measurement reference plane, with measured frequency response extending beyond \(6~\mathrm{GHz}\) \cite{julienNeitzert2024gepd}.

The S/H path additionally requires charge storage with sufficiently low droop and controlled acquisition and release behavior at cryogenic temperature.
Pauka \textit{et al.} demonstrated switched-capacitor control cells at \(100~\mathrm{mK}\), monitored a stored gate voltage over a \(30~\mathrm{min}\) hold interval, and reported leakage sufficiently low to permit refresh intervals of several minutes \cite{Pauka2021NatElectron}.
The same work reported approximately \(18~\mathrm{nW}\) average dissipation per cell for \(100~\mathrm{mV}\) control pulses.
These results provide a low-temperature precedent for charge retention over intervals far longer than the nanosecond-scale hold windows considered here.

\section{Power and Interconnect Scaling Benchmark}
\label{sec:comparative_power}

This section compares the proposed shared-template controller with full-waveform Cryo-CMOS and per-channel wavelength-division-multiplexed (WDM) photonic control. The model considers \(N\) control channels, with each channel comprising the \(I\), \(Q\), and \(Z/CZ\) resources defined in Sec.~\ref{sec:background}. We first establish common electrical-output and pulse conditions, then compare the 4~K power and RT-to-4-K interconnect scaling of the three architectures.

\subsection{Benchmark Definition and Power Model}
\label{subsec:power_model}

All three architectures are evaluated at the output of the 4~K controller, before the interconnect to the millikelvin stage. Bardin \textit{et al.} estimate approximately \(-70~\mathrm{dBm}\) of in-pulse available power at the qubit drive port for a \(50~\mathrm{MHz}\) Rabi rate. Krinner \textit{et al.} obtain approximately \(-66~\mathrm{dBm}\) peak power and \(-71~\mathrm{dBm}\) pulse-averaged power for a \(20~\mathrm{ns}\) Gaussian pulse \cite{Bardin2021Microwaves,Krinner2019EPJQT}. As a direct wiring precedent, Zanner \textit{et al.} used \(10~\mathrm{dB}\) attenuation at the still stage and \(20~\mathrm{dB}\) at the mixing chamber after the 4~K stage \cite{Zanner2022NatPhys}. We use the \(-70~\mathrm{dBm}\) value as the peak-envelope electrical reference and account for Gaussian pulse averaging separately in the optical-power calculation. With \(30~\mathrm{dB}\) between the 4~K output and the qubit, this reference requires \(-40~\mathrm{dBm}\), or \(6.33~\mathrm{mV_{pp}}\), across \(50~\Omega\) at 4~K. We select
\[
V_{\mathrm{4K,pp}}=7.4~\mathrm{mV_{pp}},
\]
which provides \(1.36~\mathrm{dB}\) of margin. The \(Z/CZ\) resource is evaluated at the same 4~K output condition; a fabricated flux-control interface would set its device-level swing through the measured line transfer function.

The optical-power calculation uses a \(20~\mathrm{ns}\) peak-normalized Gaussian centered at \(10~\mathrm{ns}\), with \(\sigma_G=3.333~\mathrm{ns}\), followed by a \(5~\mathrm{ns}\) idle interval. The corresponding derivative supplies the DRAG quadrature \cite{Motzoi2009PRL,Hyyppa2024PRXQ}. Integration over the \(25~\mathrm{ns}\) slot gives an average optical factor of \(0.333315\). The same factor is applied to each provisioned optical branch in the common benchmark. The electrical terms use their continuous simulated values.

\begin{figure*}[!t]
\centering
\includegraphics[width=0.92\textwidth]{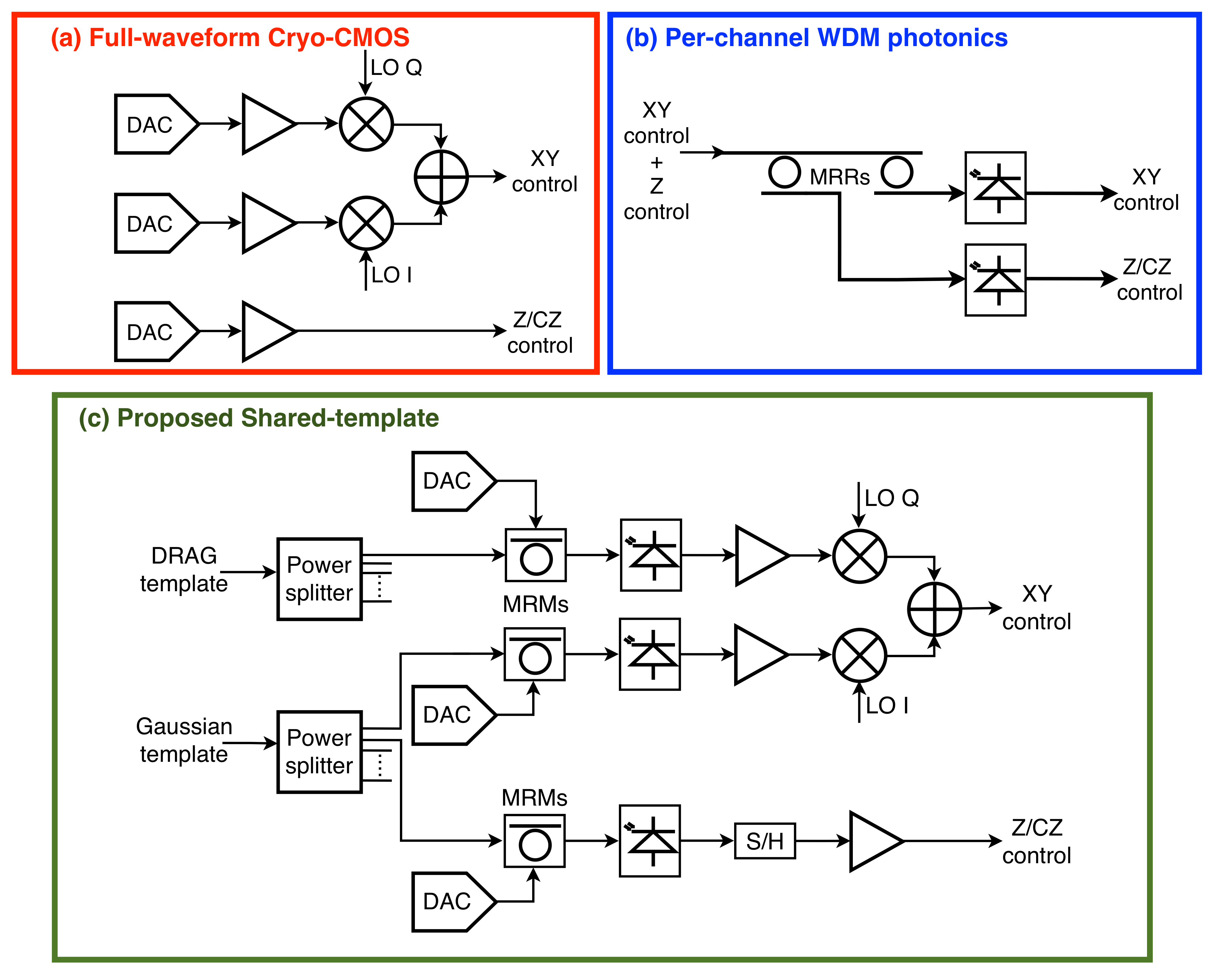}
\caption{Per-control-channel signal paths used in the power comparison. (a) Full-waveform Cryo-CMOS generates complete \(I\), \(Q\), and \(Z/CZ\) waveforms at 4~K. (b) Per-channel WDM photonics transports complete vector-\(XY\) and \(Z/CZ\) optical waveforms from room temperature. (c) The proposed controller distributes shared Gaussian and derivative templates and applies channel-specific amplitude, pulse-enable, and hold states after local optical fan-out. The three DAC symbols in (c) denote the \(I/Q/Z\) coefficient states modeled by one aggregate DAC bank.}
\label{fig:benchmark_architectures}
\end{figure*}

With the input pulse and 4~K output fixed, Fig.~\ref{fig:benchmark_architectures} shows how waveform generation and signal delivery differ among the three cases. Full-waveform Cryo-CMOS, shown in panel (a), generates complete \(I\), \(Q\), and \(Z/CZ\) waveforms locally using three high-speed DAC paths. Per-channel WDM photonics, shown in panel (b), combines the room-temperature \(I\) and \(Q\) waveforms into one vector-\(XY\) optical path and carries the complete \(Z/CZ\) waveform on a second optical path. The proposed controller, shown in panel (c), distributes shared Gaussian and derivative templates and applies channel-specific amplitude, pulse-enable, and hold states after local fan-out. Its three DAC symbols represent the programmed \(I/Q/Z\) coefficient states and share the aggregate DAC-bank power reported in Table~\ref{tab:power_model_parameters}.

The electrical terms for panels (a) and (c) come from post-layout circuit simulations using the GlobalFoundries 45SPCLO silicon-photonic/CMOS platform. The proposed power model assigns \(65.25~\mu\mathrm{W}\) to each local \(I\), \(Q\), and \(Z/CZ\) path, giving \(195.75~\mu\mathrm{W}\) per qubit-control channel. The simulated 8-bit segmented-capacitor DAC bank operates at \(100~\mathrm{MHz}\), updates the channel coefficients and control states at gate-event boundaries, and dissipates \(59.17~\mu\mathrm{W}\) per qubit-control channel. The aggregate class-AB source-follower output driver dissipates \(231.10~\mu\mathrm{W}\) per qubit-control channel \cite{RamirezAngulo2006ClassABFollower}. These contributions sum to an active electrical subtotal of \(195.75+59.17+231.10=486.02~\mu\mathrm{W}\) per qubit-control channel.

The full-waveform Cryo-CMOS reference uses the same three downstream analog-path terms and the same aggregate output-driver term as the proposed controller. Its DACs, however, serve a different function. The proposed \(100~\mathrm{MHz}\) DAC bank programs the MRM amplitude and channel state, whereas the Cryo-CMOS reference does not use an MRM. Instead, three independent 8-bit, \(2~\mathrm{GHz}\) DACs directly synthesize the complete \(I\), \(Q\), and \(Z/CZ\) waveforms at 4~K. Each DAC drives its corresponding downstream analog path, and the output-driver stage supplies the current required to reach the common \(7.4~\mathrm{mV_{pp}}\) output boundary. The selected DAC rate is consistent with demonstrated GHz-rate cryogenic envelope generation \cite{Bardin2019JSSC}, and a dedicated 45SPCLO post-layout simulation gives \(1.007~\mathrm{mW}\) per DAC. Holding the downstream analog paths and output driver fixed therefore isolates the power associated with generating three complete waveforms at \(2~\mathrm{GHz}\). The circuit simulations use the foundry device models and retain common bias and output-swing margin for expected cryogenic MOSFET shifts \cite{Beckers2020JEDS}.

Table~\ref{tab:power_model_parameters} collects the circuit-simulation results and the optical, interconnect, and sensitivity parameters used below.

\begin{table*}[!t]
\centering
\small
\renewcommand{\arraystretch}{1.12}
\caption{Electrical, optical, interconnect, and sensitivity parameters used in the power comparison.}
\label{tab:power_model_parameters}
\begin{tabular}{@{}p{0.25\textwidth} p{0.39\textwidth} p{0.28\textwidth}@{}}
\hline
\textbf{Quantity} & \textbf{Nominal value} & \textbf{Basis} \\
\hline
Reference point & \(N=1000\), \(G=16\), \(V_{\mathrm{4K,pp}}=7.4~\mathrm{mV_{pp}}\), \(Z_0=50~\Omega\) & Representative qubit-drive level and measured cold-stage attenuation configuration \cite{Bardin2021Microwaves,Krinner2019EPJQT,Zanner2022NatPhys} \\
Optical pulse & \(20~\mathrm{ns}\) Gaussian, \(5~\mathrm{ns}\) idle, \(\sigma_G=3.333~\mathrm{ns}\), slot average \(0.333315\) & Common optical-power workload; Gaussian/DRAG convention \cite{Motzoi2009PRL,Hyyppa2024PRXQ} \\
Proposed electrical circuits & \(65.25~\mu\mathrm{W}\) for each local path; S/H hold capacitor \(C_{\mathrm{H}}=214~\mathrm{fF}\); \(231.10~\mu\mathrm{W}\) aggregate driver; 8-bit, \(100~\mathrm{MHz}\), \(59.17~\mu\mathrm{W}\) aggregate DAC bank; \(C_{\mathrm{unit}}=31.2~\mathrm{fF}\) & Circuit-simulation-based local-path, driver, and DAC values; capacitor values included in the corresponding simulations \cite{RamirezAngulo2006ClassABFollower} \\
Full-waveform Cryo-CMOS & Three 8-bit, \(2~\mathrm{GHz}\) DACs at \(1.007~\mathrm{mW/instance}\) & Extracted DAC result with matched downstream paths; DAC-rate basis from \cite{Bardin2019JSSC} \\
Optical conversion & \(R_{\mathrm{PD}}=0.1~\mathrm{A/W}\); proposed \(I_{\mathrm{branch,pk}}=12~\mu\mathrm{A}\) & Common detector responsivity; proposed current from the extracted receiver path \cite{julienNeitzert2024gepd} \\
Optical losses & Package loss \(3.0~\mathrm{dB}\); WDM selection/routing loss \(3.0~\mathrm{dB}\); proposed routing loss \(0.5~\mathrm{dB}\), fixed insertion loss \(0.9~\mathrm{dB}\), MMI excess loss \(0.2~\mathrm{dB/stage}\), I/Z splitting loss \(0.2~\mathrm{dB}\) & Cryogenic package, WDM, and splitter benchmarks; WDM loss is an engineering allowance \cite{Lin2023CryoPWB,Zeng2023CryoPackaging,Fard2020CryoWDM,Bisang2026CryoWDM,Yi2023Splitter,Sheng2012MMI} \\
LO and passive interconnect & \(4.243~\mu\mathrm{W/channel}\); \(5.6~\mu\mathrm{W/fiber}\); four LO coaxes at \(1.0~\mathrm{mW/coax}\); nominal 16 WDM wavelengths/fiber & Local LO and thermal model \cite{Joshi2024JLT}; WDM packing is a benchmark assumption \\
Shared sensitivity ranges & \(R_{\mathrm{PD}}=0.08\)--\(0.125~\mathrm{A/W}\); package loss \(2.5\)--\(3.5~\mathrm{dB}\) & Shared photonic ranges \cite{julienNeitzert2024gepd,Lin2023CryoPWB,Zeng2023CryoPackaging} \\
Proposed sensitivity ranges & Electrical terms \(\pm7\%\); routing \(0.2\)--\(1.0~\mathrm{dB}\); MMI \(0.05\)--\(0.40~\mathrm{dB/stage}\); I/Z split \(0.05\)--\(0.50~\mathrm{dB}\); LO current \(\pm10\%\); LO excess \(0.3\)--\(2.6~\mathrm{dB}\); coax \(0.8\)--\(1.2~\mathrm{mW/coax}\) & Proposed engineering sensitivity range; fixed insertion held at \(0.9~\mathrm{dB}\) \\
WDM sensitivity ranges & Selection/routing loss \(2.5\)--\(3.5~\mathrm{dB}\); fiber heat \(4.5\)--\(6.5~\mu\mathrm{W/fiber}\) & WDM engineering sensitivity range \cite{Fard2020CryoWDM,Bisang2026CryoWDM,Joshi2024JLT} \\
\hline
\end{tabular}
\end{table*}

The optical-power comparison uses a fixed \(25~\mathrm{ns}\) reference slot. In the proposed architecture, the Gaussian and derivative templates are distributed before per-channel pulse blocking; the launch-power model therefore provisions the \(I\), \(Q\), and \(Z/CZ\) endpoint branches in every slot, including channels that subsequently block the pulse. For the WDM reference, the model counts one active complete waveform, either vector-\(XY\) or \(Z/CZ\), per qubit-control channel and slot, while both installed optical paths contribute to the passive fiber heat.

The two photonic cases use the same Ge-photodetector responsivity and fiber-to-chip package loss. The proposed detector current is set by the extracted local receiver path, after which the class-AB stage supplies the current required at the common 4~K output. The WDM photodetector directly supplies that output current. The WDM path also includes the MRR wavelength-selection and routing loss listed in Table~\ref{tab:power_model_parameters}. The proposed path includes local routing, fixed device insertion, Gaussian \(I/Z\) splitting, and multimode-interference (MMI) fan-out loss.

For a local group of size \(G\), the excess loss of the MMI tree is
\begin{equation}
L_{\mathrm{split,ex}}(G)
=
\left\lceil\log_2G\right\rceil
\ell_{\mathrm{split}},
\label{eq:splitter_excess}
\end{equation}
where \(\ell_{\mathrm{split}}=0.2~\mathrm{dB/stage}\) \cite{Yi2023Splitter,Sheng2012MMI}. The \(G=16\) baseline has four splitter stages and \(0.8~\mathrm{dB}\) of excess loss. Endpoint multiplicity accounts for ideal optical division, and the final partial group is evaluated using its actual occupancy.

The number of local fan-out groups is \(\lceil N/G\rceil\). Each group receives one Gaussian-template fiber and one derivative-template fiber, giving
\begin{equation}
N_{\mathrm{fiber,prop}}
=
2\left\lceil\frac{N}{G}\right\rceil .
\label{eq:proposed_fiber_count}
\end{equation}
The passive heat model uses \(5.6~\mu\mathrm{W}\) per optical fiber \cite{Joshi2024JLT}. The proposed microwave path also uses four shared \(6~\mathrm{GHz}\) LO frequency-class feeds. Mixer loading and passive RF-distribution excess give the per-channel LO term in Table~\ref{tab:power_model_parameters}, while the four coaxial feeds provide a fixed RT-to-4-K thermal load.

The sensitivity analysis evaluates 3500 parameter sets. Photodetector responsivity is sampled logarithmically, and the other quantities are sampled uniformly over the ranges in Table~\ref{tab:power_model_parameters}. Each photonic trial uses the same responsivity and package-loss sample for the proposed and WDM cases, and one common electrical scale factor is applied to the proposed local paths, output driver, and DAC bank. At each \(N\), the larger deviation of the 10th and 90th percentiles from the nominal result defines the plotted sensitivity band. Because the input intervals are engineering ranges rather than measured process distributions, the bands indicate sensitivity to those ranges and should not be interpreted as statistical confidence intervals or yield predictions. The Cryo-CMOS curve uses the fixed comparison values and is plotted without a sensitivity band.

The total 4~K power combines the active electrical circuits, optical signal delivery, local LO loading, and passive interconnect heat. Most channel circuits scale linearly with \(N\); the four LO coaxes add a fixed load, and the template-fiber term changes in steps with \(\lceil N/G\rceil\).

\subsection{Modeled Power Results and Interconnect Trade-Offs}
\label{subsec:scaling_results}

\begin{figure}[t]
\centering
\includegraphics[width=\columnwidth]{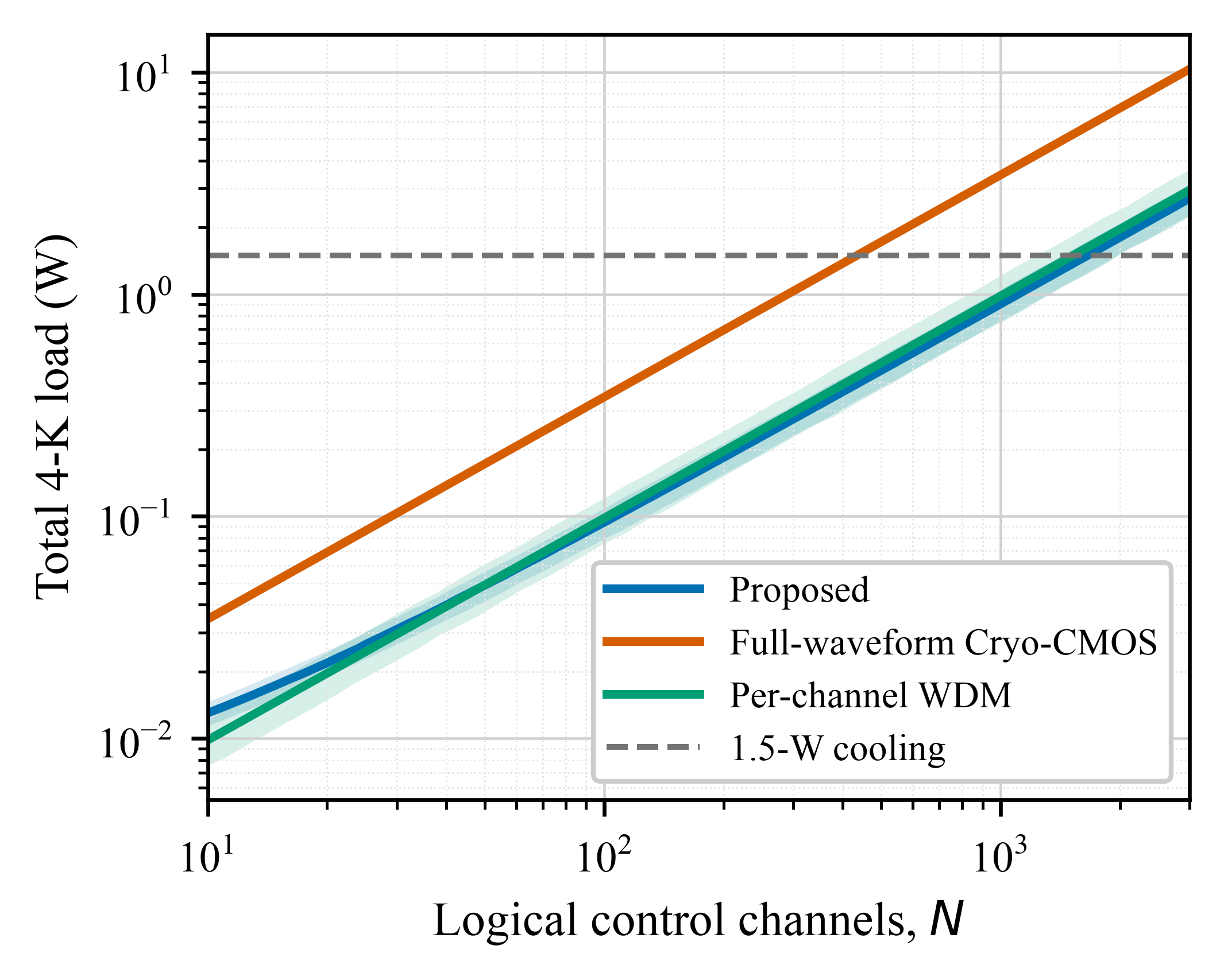}
\caption{Modeled total 4~K control-downlink power. The proposed case uses \(G=16\), includes the three local \(I\), \(Q\), and \(Z/CZ\) paths defined in the benchmark, and uses two optical-template fibers per local group plus four shared LO feeds. The dashed line marks a representative \(1.5~\mathrm{W}\) 4~K stage capacity \cite{Krinner2019EPJQT}. Shaded regions show the engineering sensitivity ranges of the two photonic cases.}
\label{fig:power_scaling}
\end{figure}

\begin{figure}[t]
\centering
\includegraphics[width=\columnwidth]{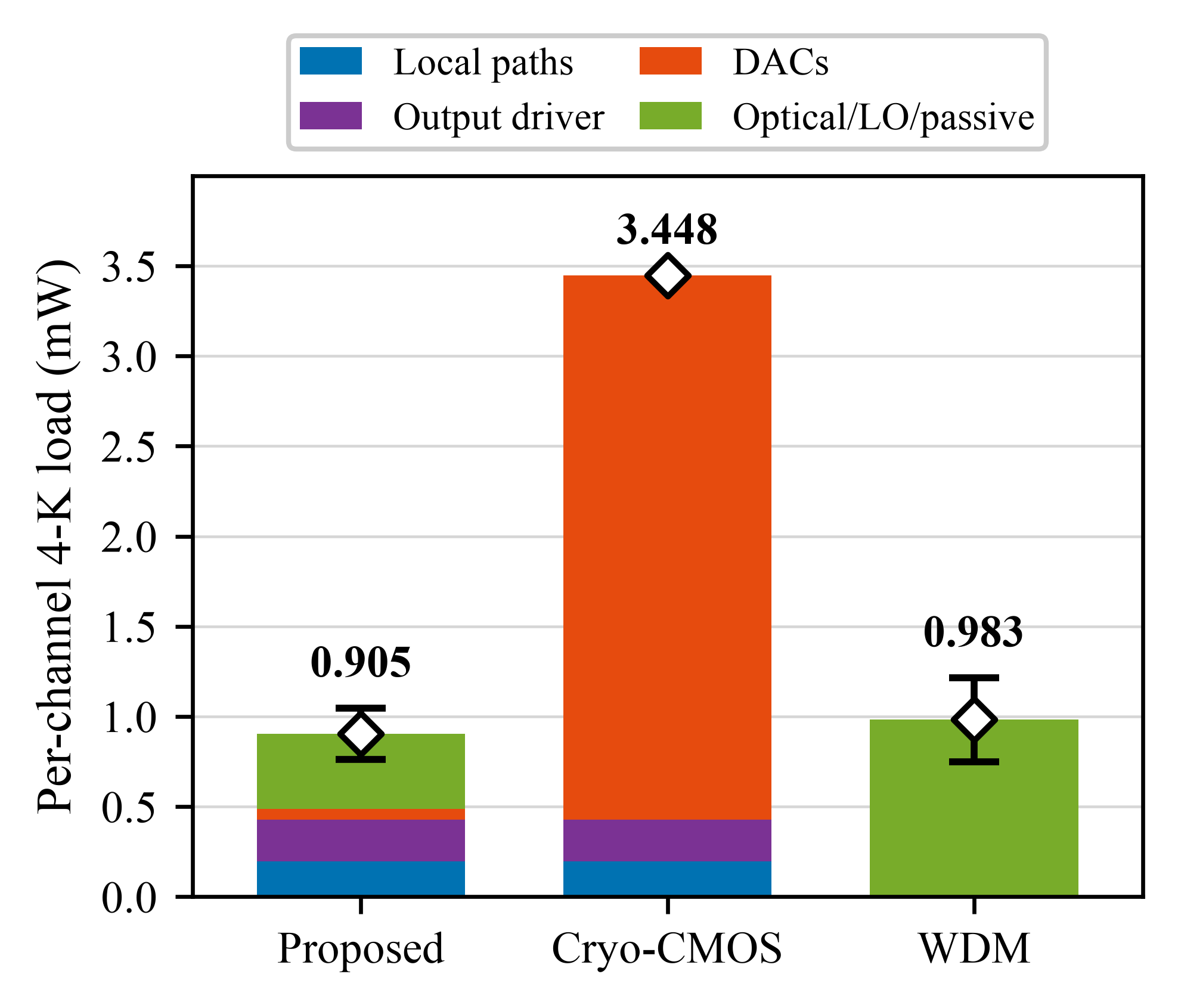}
\caption{Modeled 4~K power composition per qubit-control channel at \(N=1000\). The stacked bars separate the local paths, aggregate output driver, DACs, and combined optical, LO, and passive-interconnect contributions. Capped bars show the sensitivity ranges of the two photonic cases, and diamonds mark their nominal values.}
\label{fig:power_breakdown}
\end{figure}

Fig.~\ref{fig:power_scaling} shows the total 4~K control-downlink power as the number of qubit-control channels increases, and Fig.~\ref{fig:power_breakdown} resolves the totals at \(N=1000\) into their main contributions.

At \(N=1000\), the modeled power is \(0.905~\mathrm{mW}\) per qubit-control channel for the proposed controller, \(3.448~\mathrm{mW}\) for the normalized full-waveform Cryo-CMOS reference, and \(0.983~\mathrm{mW}\) for the direct-detection WDM reference. Under the common \(50~\Omega\) output condition, the proposed value is approximately \(74\%\) below the Cryo-CMOS value and \(8\%\) below the nominal WDM value. The breakdown below explains these differences and why the ordering of the two photonic cases depends on the receiver design.

In the proposed controller, optical signal delivery contributes \(0.410~\mathrm{mW}\) per qubit-control channel, or approximately \(45\%\) of the total. The aggregate output driver contributes \(0.231~\mathrm{mW}\), the three local paths contribute \(0.196~\mathrm{mW}\), and the coefficient-DAC bank contributes \(0.0592~\mathrm{mW}\). The local LO and passive-interconnect terms together contribute approximately \(0.009~\mathrm{mW}\). In the full-waveform Cryo-CMOS reference, the three waveform DACs contribute \(3.021~\mathrm{mW}\) and dominate the total. Moving complete waveform generation out of the 4~K stage therefore removes most of the GHz-rate DAC power; optical delivery and the output driver then become the largest terms in the proposed controller.

The nominal WDM result is set mainly by its optical-to-electrical conversion boundary. The WDM photodetector directly supplies the current required to produce \(7.4~\mathrm{mV_{pp}}\) across \(50~\Omega\), so the adopted detector responsivity together with the package and wavelength-selection losses produces an optical contribution of \(0.982~\mathrm{mW}\) per qubit-control channel. This value is an O-band benchmark based on the common \(0.1~\mathrm{A/W}\) detector assumption defined above. In the proposed controller, the photodetector recovers a lower-current envelope and the class-AB stage supplies the final output current; the power of that driver is already included in the proposed total. This difference makes the nominal direct-detection WDM result slightly higher, while the shared dependence on detector responsivity and package loss causes the two photonic sensitivity ranges to overlap.

The receiver impedance changes this comparison substantially. Repeating the same WDM calculation with an ideal lossless 4:1 impedance transformation allows the detector to see \(200~\Omega\) while preserving the common \(50~\Omega\) output boundary. The required photocurrent then decreases by a factor of two, giving a modeled \(0.492~\mathrm{mW}\) per qubit-control channel, approximately \(46\%\) below the proposed nominal value. A practical matching network must preserve the required bandwidth in the presence of detector capacitance and will introduce finite insertion loss. A post-detection driver makes a related trade by replacing part of the detector optical-power requirement with active electrical power. A WDM receiver that realizes either option with sufficiently low overhead can dissipate less power than the proposed controller because it avoids the proposed local template-routing and amplitude-control circuitry.

The fixed shared resources determine the low-channel-count behavior in Fig.~\ref{fig:power_scaling}. At small \(N\), the four LO coaxes and the first pair of template fibers are shared by only a few channels, placing the proposed curve above WDM. The nominal curves cross near \(N=50\) as these fixed loads are amortized. The small steps in the proposed curve follow from the integer number of local groups, \(\lceil N/G\rceil\), and from the smaller splitter tree used by a partially populated final group.

At \(N=1000\), the proposed control-downlink dissipates \(0.905~\mathrm{W}\), or approximately \(60\%\) of the representative \(1.5~\mathrm{W}\) 4~K cooling capacity. For this representative cooling capacity, \(0.595~\mathrm{W}\) would remain for the other 4~K functions, including readout, digital control, photonic tuning, package infrastructure, and operating margin. The modeled reduction therefore relieves one important 4~K load, while the complete fault-tolerant system budget still depends on the remaining control, readout, and cryogenic-computing subsystems.

\begin{figure}[t]
\centering
\includegraphics[width=\columnwidth]{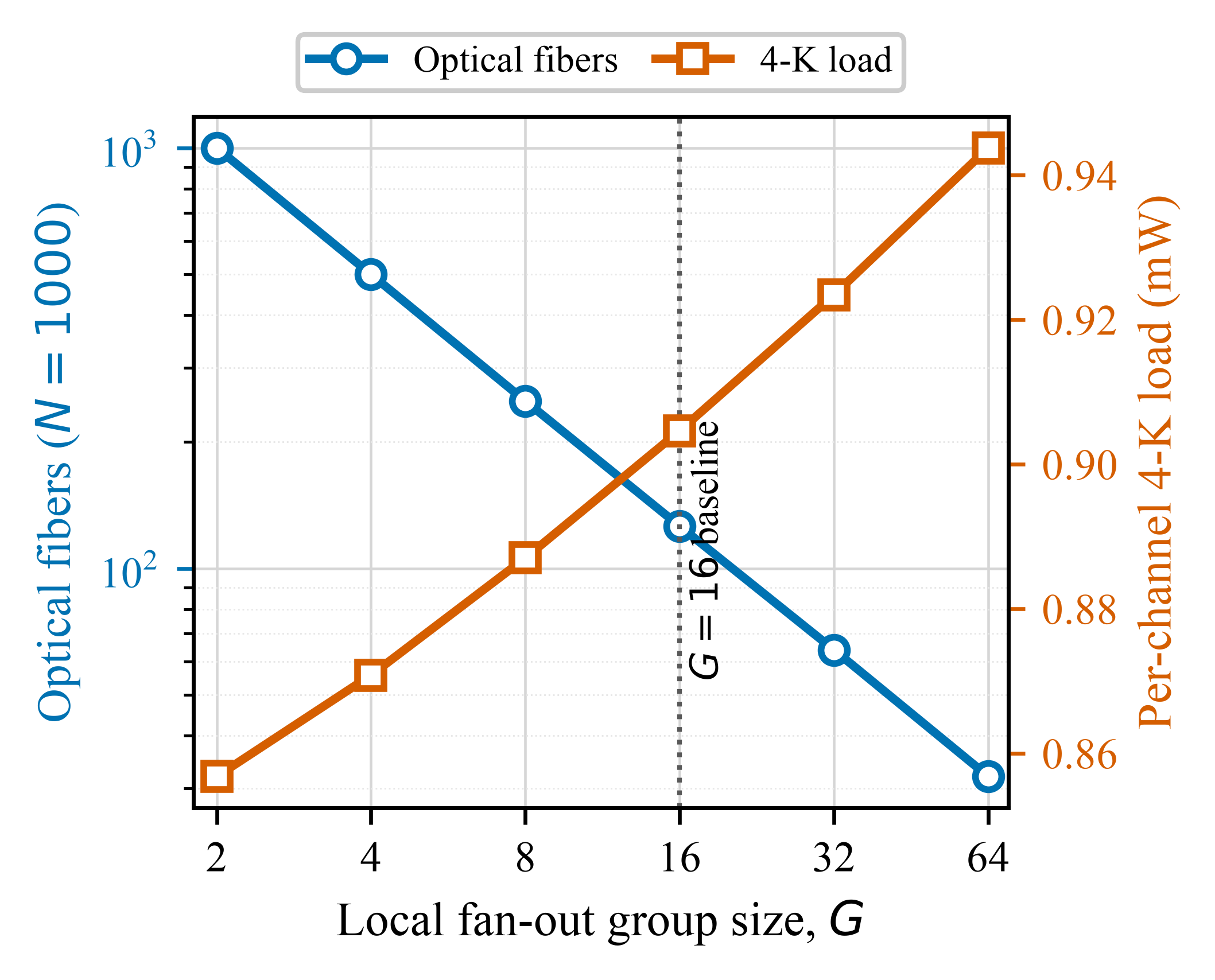}
\caption{Trade-off between local fan-out size, RT-to-4-K optical-template fiber count, and modeled 4~K power per qubit-control channel at \(N=1000\). Increasing \(G\) reduces the number of spatial template fibers while increasing the MMI depth and optical loss. The fiber curve counts the Gaussian and derivative feeds; the power curve also includes the fixed four-coax LO load.}
\label{fig:fanout_tradeoff}
\end{figure}

Fig.~\ref{fig:fanout_tradeoff} shows how the proposed optical power and interconnect count change with local fan-out.

Increasing \(G\) allows more channels to share each pair of template fibers, but every doubling of a full group adds one MMI stage and its associated excess loss. We select \(G=16\) because it is the smallest power-of-two group size that brings the proposed fiber count close to the nominal WDM packing at \(N=1000\): \(63\) groups require \(126\) template fibers, compared with \(125\) WDM fibers for \(W=16\) and two installed optical waveform paths per qubit-control channel. Smaller groups reduce the proposed optical power but require substantially more fibers, whereas larger groups save fibers at the cost of additional splitter depth and launch power.

The thermal and packaging consequences of these connections are different. Under the adopted model, one optical fiber conducts \(5.6~\mu\mathrm{W}\) from room temperature to 4~K, compared with \(1.0~\mathrm{mW}\) for one LO coaxial feed \cite{Joshi2024JLT}. For the proposed \(G=16\) case, the modeled template-and-LO downlink comprises 126 optical-template fibers and four shared LO coaxial feeds. The fibers dominate the connection count, while the coaxes dominate the passive conductive load. The large fiber count primarily affects package density, assembly, and reliability, and contributes only a small fraction of the modeled 4~K power.

The comparison with \(125\) WDM fibers depends on the nominal packing factor \(W=16\). Wavelength multiplexing avoids the spatial-division loss of the proposed fan-out network, but increasing \(W\) shifts the scaling constraint to the wavelength-selective receiver. The available optical bandwidth and resonator free spectral range, filter linewidth and channel spacing, accumulated insertion loss and crosstalk, and resonance alignment determine how many channels can be carried by one fiber \cite{Yuan2024MRM,Fard2020CryoWDM,adya2025nvRing}. Cryogenic experiments have demonstrated a four-channel WDM photonic circuit at \(40~\mathrm{K}\) and an eight-channel arrayed-waveguide grating based optical-to-microwave receiver at \(4~\mathrm{K}\), with two wavelengths operated simultaneously in each case \cite{Fard2020CryoWDM,Bisang2026CryoWDM}. In the present model, \(W=16\) is used as a system-level packing target. Reaching this target requires all 16 receiver channels to satisfy the assumed loss, crosstalk, and wavelength-alignment conditions simultaneously. The proposed and WDM approaches therefore exchange fiber count for on-chip photonic complexity through spatial fan-out and wavelength selection, respectively.

These results expose three different system trade-offs. Full-waveform Cryo-CMOS minimizes high-speed room-temperature-to-4~K waveform transport but pays the largest modeled active-power cost for local GHz-rate waveform generation. An efficiently matched or driver-assisted WDM receiver can provide lower photonic power and denser fiber packing, while complete channel-specific waveforms are generated and assigned before optical delivery. The proposed controller uses more spatial optical transport and local analog circuitry in exchange for amplitude, pulse blocking, and hold control after template distribution, without a per-channel GHz-rate waveform generator. Improved photodetector responsivity, lower optical loss, and a more efficient output driver provide the largest opportunities for reducing its power further.

\section{Control-Chain Fidelity Analysis}
\label{sec:fidelity}

\subsection{Simulation Method and Error Metrics}
\label{subsec:fidelity_considerations}

This section maps the simulated controller outputs to closed-system gate errors. The separately simulated \(I\)- and \(Q\)-path transients drive a five-level single-transmon model, and the simulated S/H output drives a coupled-transmon conditional-phase model. QuTiP integrates each time-dependent Hamiltonian, after which the propagated operation is evaluated over computational-subspace inputs \cite{qutip5}. The selected hardware perturbations follow the carrier/envelope decomposition used for controller specifications \cite{vanDijk2019PRApplied}, while the single-qubit model follows Gaussian-plus-derivative DRAG control \cite{Motzoi2009PRL,Hyyppa2024PRXQ}.

Let \(U_{\lambda}\) be the propagated operation for case \(\lambda\), and let \(V\) embed the \(d\)-dimensional computational basis into the full Hilbert space. The projected block is \(M_{\lambda}=V^{\dagger}U_{\lambda}V\). For comparison with a logical target, \(B_{\lambda}=U_{\mathrm{tar}}^{\dagger}M_{\lambda}\); for comparison with a propagated reference, \(B_{\lambda}=V^{\dagger}U_{\mathrm{ref}}^{\dagger}U_{\lambda}V\). The error and final-leakage metrics are
\begin{align}
\epsilon_{\mathrm{avg}}(\lambda)
&=1-\frac{\operatorname{Tr}(B_{\lambda}^{\dagger}B_{\lambda})
+\left|\operatorname{Tr}B_{\lambda}\right|^2}{d(d+1)},
\label{eq:average_control_error}\\
\epsilon_{\mathrm{ws}}(\lambda)
&=1-\min_{\substack{|\psi\rangle\in\mathbb{C}^{d}\\
\langle\psi|\psi\rangle=1}}
\left|\langle\psi|B_{\lambda}|\psi\rangle\right|^2,
\label{eq:worst_state_control_error}\\
L_{\mathrm{avg}}(\lambda)
&=1-\frac{\operatorname{Tr}(M_{\lambda}^{\dagger}M_{\lambda})}{d},
\qquad
L_{\mathrm{max}}(\lambda)
=1-\lambda_{\min}(M_{\lambda}^{\dagger}M_{\lambda}).
\label{eq:population_leakage_metrics}
\end{align}
Here, \(d=2\) for the single-qubit model and \(d=4\) for the two-qubit model. The quantities \(\epsilon_{\mathrm{avg}}\) and \(\epsilon_{\mathrm{ws}}\) are the average and largest overlap errors over computational inputs, while \(L_{\mathrm{avg}}\) and \(L_{\mathrm{max}}\) are the corresponding average and largest final population loss. Target-relative overlap includes absolute leakage, while leakage shared by two reference-relative evolutions can cancel and is therefore reported separately.

The simulated single-qubit pulses are compared with the calibrated DRAG operation, additional addressed-gate perturbations are compared with simulated pulse~1, and unwanted-drive cases are compared with zero-drive evolution. The \(10^{-3}\) line marks the common plotted error scale. The single-qubit figures use \(\epsilon_{\mathrm{ws}}\), and the two-qubit figure uses \(\epsilon_{\mathrm{avg}}\).

\subsection{Single-Qubit XY/DRAG Results}
\label{subsec:single_qubit_sensitivity}

The single-qubit calculation retains five levels of a Duffing transmon. In the drive rotating frame,
\begin{equation}
\begin{aligned}
\frac{H_{\mathrm{RWA}}(t)}{\hbar}={}&
\Delta_{\mathrm{cal}}n
+\frac{\alpha}{2}n(n-1)\\
&+\frac{\Omega_I(t)}{2}(a+a^{\dagger})
+\frac{\Omega_Q(t)}{2}\left[-i(a-a^{\dagger})\right],
\end{aligned}
\label{eq:five_level_drag_model}
\end{equation}
where \(n=a^{\dagger}a\), \(\Delta_{\mathrm{cal}}=\omega_{01}-\omega_d\), \(\alpha/2\pi=-250~\mathrm{MHz}\), and \(\Omega_I\) and \(\Omega_Q\) are the two Rabi-rate quadratures. For a \(20~\mathrm{ns}\) pulse centered at \(t_c=10~\mathrm{ns}\),
\begin{equation}
\begin{aligned}
g_G(t)&=\exp\!\left[-\frac{(t-t_c)^2}{2\sigma_G^2}\right],
&q_G(t)&=-\frac{t-t_c}{\sigma_G^2}g_G(t),\\
\Omega_I(t)&=\Omega_{\mathrm{pk}}g_G(t),
&\Omega_Q(t)&=\Omega_{\mathrm{pk}}\beta_D q_G(t),
\end{aligned}
\label{eq:drag_basis_definition}
\end{equation}
with \(g_G=q_G=0\) outside the gate window and \(\sigma_G=2.66~\mathrm{ns}\). Joint calibration gives \(\Omega_{\mathrm{pk}}/2\pi=70.623~\mathrm{MHz}\), \(\beta_D=1.07788~\mathrm{ns}\), and \(\Delta_{\mathrm{cal}}/2\pi=15.509~\mathrm{MHz}\), with an average overlap error of \(2.18\times10^{-7}\) relative to an ideal \(X_{\pi}\). The optical-power workload in Sec.~\ref{sec:comparative_power} uses its separately defined \(\sigma_G=3.333~\mathrm{ns}\).

A preliminary circuit run fixes pulse timing and Q-path predistortion. The analyzed \(I\)- and \(Q\)-path transients are then coherently demodulated at \(6~\mathrm{GHz}\), baseline corrected, and mapped to Rabi-rate units by assigning the \(7.4~\mathrm{mV}_{\mathrm{pp}}\) controller reference to the calibrated \(\Omega_{\mathrm{pk}}\). In the gate model, the two scalar paths form orthogonal mixer quadratures and the known derivative sign restores signed \(Q\). Pulse~1 calibrates two path coefficients and one virtual-\(Z\) phase. Their 8-bit values are codes 232 and 59, corresponding to \(c_I=0.90980\) and \(c_Q=0.23137\). Pulse~2 occurs \(100~\mathrm{ns}\) later in the same record and uses these values unchanged. The behaviorally combined pulse-1 envelope is \(7.384~\mathrm{mV}_{\mathrm{pp}}\).

Pulse~1 gives \((\epsilon_{\mathrm{avg}},\epsilon_{\mathrm{ws}})=(1.67\times10^{-5},3.20\times10^{-5})\), and pulse~2 gives \((2.84\times10^{-4},4.32\times10^{-4})\). The pulse-2 \(I\)- and \(Q\)-path peaks are \(0.9856\) and \(0.9796\) of their pulse-1 values, consistent with the larger transferred-calibration error. Maximum final leakage remains below \(2.8\times10^{-5}\) for both pulses.

The remaining cases perturb one parameter at a time. They include a \(1\%\) common gain change, a \(\pm100~\mathrm{ps}\) Q-to-I delay, and a \(100~\mathrm{ps}\) common-width test implemented as a \(\pm0.5\%\) time scaling about the pulse center, with fixed peak amplitude and a fixed \(0\)--\(20~\mathrm{ns}\) propagation window. The \(5.18~\mathrm{mrad}\) LO-phase case represents a one-standard-deviation-equivalent quasistatic offset derived from reported \(137.4~\mathrm{fs}\) integrated jitter at \(6~\mathrm{GHz}\) \cite{norimatsu2025rfg}. The other cases use a \(0.13\%\) photodetector pulse-area change, \(20~\mathrm{dB}\) end-to-end optical extinction, \(59.7~\mathrm{dBc}\) LO feedthrough relative to the calibrated peak drive over \(20~\mathrm{ns}\), and \(35~\mathrm{dB}\) spectator-tone suppression at \(+100~\mathrm{MHz}\) \cite{norimatsu2025rfg,acharya2023mux}. Direct photodetection maps optical-power extinction to an envelope ratio \(10^{-ER/10}\), while RF-power suppression maps to a field ratio \(10^{-D/20}\).

\begin{figure}[t]
\centering
\includegraphics[width=\columnwidth]{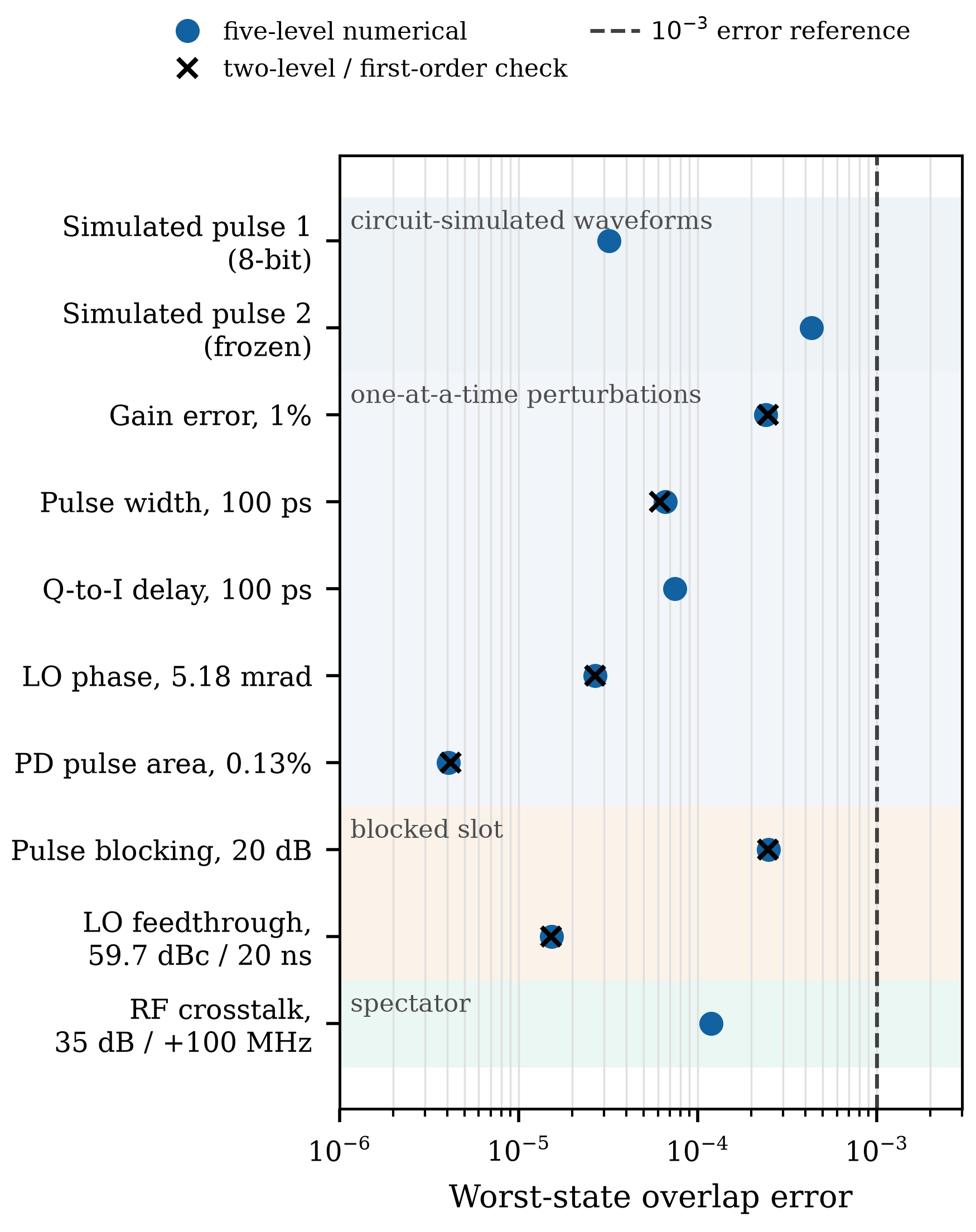}
\caption{Numerically estimated worst-state overlap errors for representative single-qubit cases in the five-level model. The first two rows compare the simulated I/Q waveforms with the calibrated DRAG operation; the pulse-1 calibration is held fixed for pulse~2. The addressed-gate perturbations are evaluated relative to pulse~1, while pulse blocking, LO feedthrough, and spectator crosstalk are evaluated relative to zero-drive evolution. Blue circles show five-level results, black crosses show two-level or first-order checks, and the dashed line is the \(10^{-3}\) error reference.}
\label{fig:controller_error_terms}
\end{figure}

All cases in Fig.~\ref{fig:controller_error_terms} remain below the \(10^{-3}\) worst-state error reference. The \(1\%\) common-gain case gives \(2.41\times10^{-4}\) relative to pulse~1. The \(20~\mathrm{dB}\) pulse-blocking and \(35~\mathrm{dB}\) spectator cases give \(2.50\times10^{-4}\) and \(1.19\times10^{-4}\), respectively, while the other imposed perturbations remain below \(7.5\times10^{-5}\).

\begin{figure*}[t]
\centering
\includegraphics[width=\textwidth]{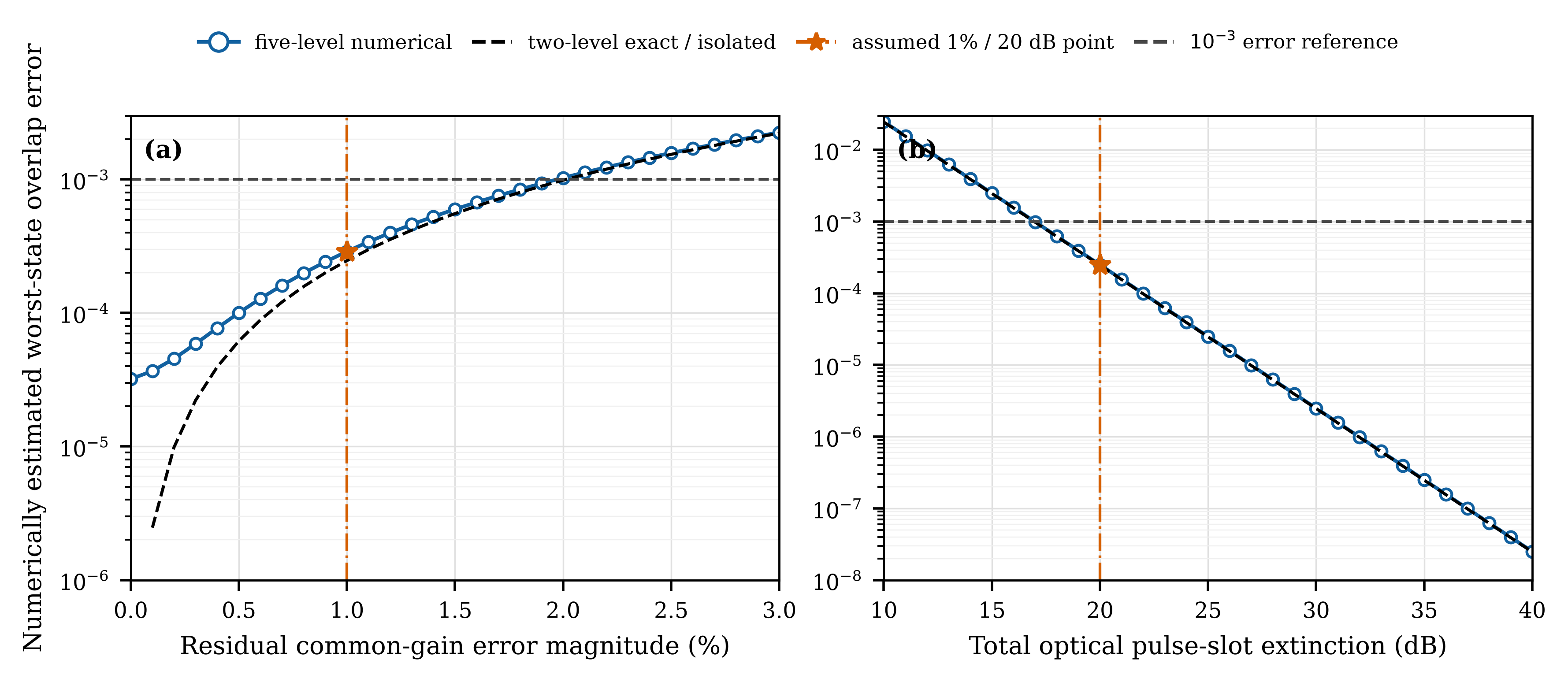}
\caption{Selected single-qubit sensitivity sweeps. (a) Worst-state overlap error relative to the calibrated DRAG operation versus residual common-gain error. The zero-error point retains the simulated-waveform and 8-bit coefficient mismatch. (b) Blocked-slot overlap error relative to zero-drive evolution versus total optical-power extinction at the photodetector input. Blue markers show five-level results, black dashed curves show isolated two-level checks, orange stars mark the \(1\%\) and \(20~\mathrm{dB}\) cases, and the horizontal dashed lines show the \(10^{-3}\) error reference.}
\label{fig:selected_controller_sweeps}
\end{figure*}

Fig.~\ref{fig:selected_controller_sweeps}(a) gives a total worst-state error of \(2.89\times10^{-4}\) at \(1\%\) gain error, including the baseline waveform and quantization mismatch, and reaches \(10^{-3}\) near \(1.97\%\). In panel~(b), \(20~\mathrm{dB}\) end-to-end extinction gives \(2.50\times10^{-4}\), and the crossing occurs near \(17.0~\mathrm{dB}\). The relevant device quantity is the extinction of the complete blocking path at the photodetector input \cite{gevorgyan2021cryoRing}.

\subsection{Baseband Two-Qubit Envelope Results}
\label{subsec:two_qubit_envelope_sensitivity}

The two-qubit input is the simulated S/H output for a smooth-edged envelope with a \(10~\mathrm{ns}\) rise, an \(80~\mathrm{ns}\) hold, and a \(10~\mathrm{ns}\) fall. The ideal edges are endpoint-normalized truncated Gaussians with \(\sigma_G=2.66~\mathrm{ns}\); the envelope is active from \(20\) to \(120~\mathrm{ns}\), and both propagation and pulse-area comparison use the \(0\)--\(150~\mathrm{ns}\) window. After baseline subtraction and flat-top normalization, the simulated waveform has \(0.476\%\) peak-to-peak flat-top variation and a \(-0.087\%\) pulse-area difference from the ideal envelope. At the end of the hold interval, the reset switch releases the stored charge through the discharge path. The simulated waveform contains a \(0.51\%\) turn-off transient followed by an RC-like discharge tail with a fitted \(1.91~\mathrm{ns}\) time constant. Both features modify the pulse area and the return trajectory toward the idle point, so the complete simulated falling edge is retained in the two-qubit propagation.

The waveform drives a representative pair of directly coupled transmons \cite{Barends2014Nature,Li2024PRX}:
\begin{equation}
\frac{H(t)}{\hbar}
=\sum_{j=1}^{2}\left[
\Delta_j(t)n_j+\frac{\alpha_j}{2}n_j(n_j-1)
\right]
+J\left(a_1^{\dagger}a_2+a_1a_2^{\dagger}\right),
\label{eq:two_qubit_envelope_model}
\end{equation}
where \(n_j=a_j^{\dagger}a_j\), \(\Delta_j\) is the rotating-frame detuning, \(\alpha_j\) is the anharmonicity, and \(J\) is the exchange coupling. The model uses three levels per transmon, \(\alpha_1/2\pi=-250~\mathrm{MHz}\), \(\alpha_2/2\pi=-300~\mathrm{MHz}\), and \(J/2\pi=20~\mathrm{MHz}\). In the qubit-2 rotating frame, \(\Delta_2=0\) and
\begin{equation}
\frac{\Delta_1(t)}{2\pi}
=-600~\mathrm{MHz}+A_f u(t),
\label{eq:two_qubit_control_mapping}
\end{equation}
where \(u(t)\) is the normalized S/H envelope and \(A_f\) is the frequency-excursion amplitude. These parameters give a bare idle \(|02\rangle\)--\(|11\rangle\) separation of \(300~\mathrm{MHz}\). The computational basis uses the four dressed idle states with maximum unique overlap with \(|00\rangle\), \(|01\rangle\), \(|10\rangle\), and \(|11\rangle\). Equal-duration idle evolution is removed before projection, and the resulting block is compared with \(\operatorname{diag}(1,1,1,-1)\) after optimizing two output-side local-\(Z\) phases.

The ideal-envelope calibration scans \(A_f=96\)--\(342~\mathrm{MHz}\). The local-phase-invariant conditional phase \(\arg[M_{00,00}M_{11,11}/(M_{01,01}M_{10,10})]\) is continuously unwrapped across this scan. Among the resulting odd-\(\pi\) branches, the lowest ideal-envelope average final leakage occurs at \(A_f=281.083~\mathrm{MHz}\) on the \(3\pi\) branch, which is logically equivalent to \(\pi\) modulo \(2\pi\).

\begin{figure*}[t]
\centering
\includegraphics[width=\textwidth]{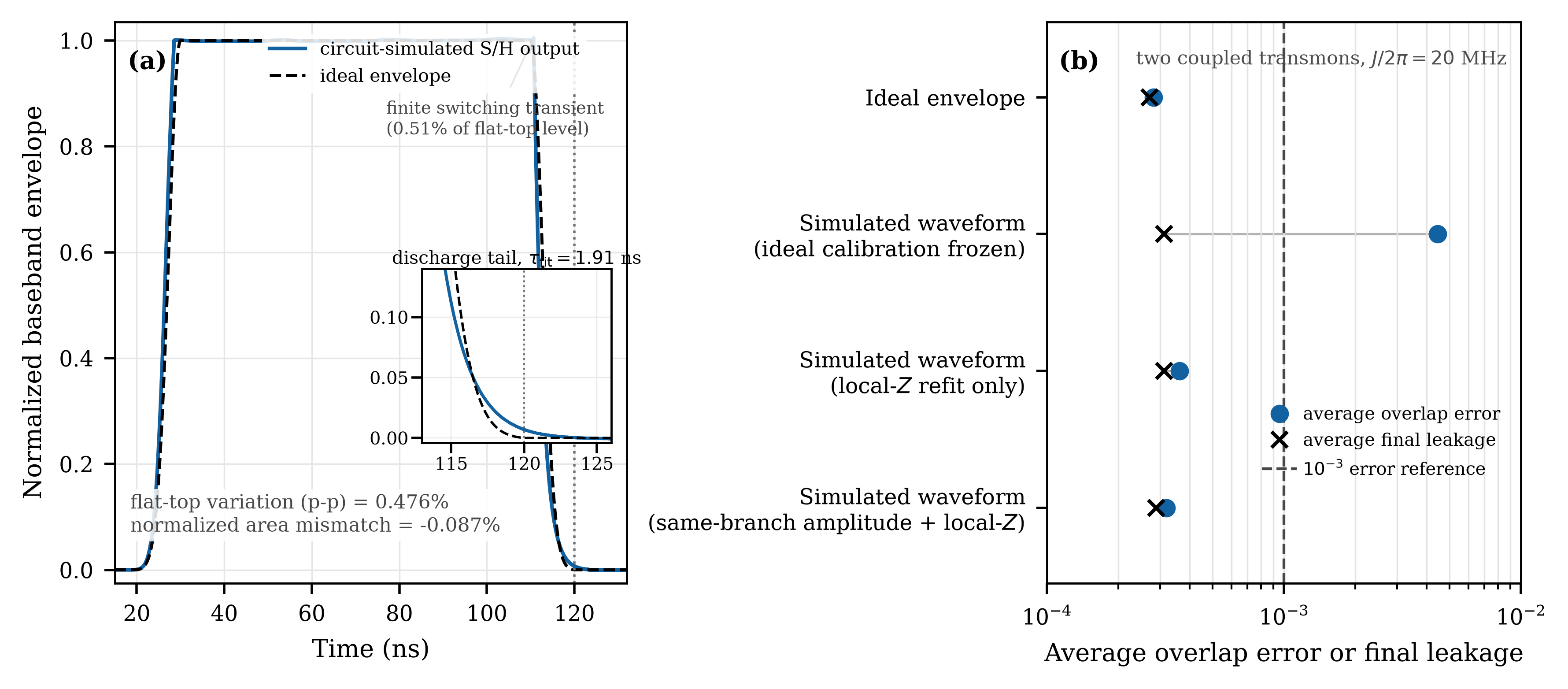}
\caption{Sensitivity of a representative baseband conditional-phase operation to the simulated S/H waveform. (a) Baseline-subtracted and flat-top-normalized S/H output compared with the ideal \(10/80/10~\mathrm{ns}\) envelope. The inset shows the falling-edge transient and discharge. (b) Computational-input average overlap error and average final leakage in the coupled-transmon model. The first simulated-waveform row transfers the ideal excursion and local-\(Z\) phases unchanged; the second refits the two local-\(Z\) phases; the third recalibrates one excursion amplitude on the same \(3\pi\) branch together with the local-\(Z\) phases. The dashed line is the \(10^{-3}\) reference for the average overlap error.}
\label{fig:two_qubit_envelope_sensitivity}
\end{figure*}

Fig.~\ref{fig:two_qubit_envelope_sensitivity}(a) compares the simulated S/H output with the ideal square-Gaussian envelope and enlarges the falling-edge transient and discharge tail. Panel~(b) shows the corresponding coupled-transmon results. The calibrated ideal envelope gives an average overlap error of \(2.81\times10^{-4}\). Applying the same frequency excursion and local-\(Z\) phases to the simulated waveform increases the error to \(4.45\times10^{-3}\). Refitting only the local-\(Z\) phases reduces it to \(3.62\times10^{-4}\), indicating that most of the transferred-calibration penalty arises from correctable single-qubit phase shifts. Recalibrating the frequency excursion on the same \(3\pi\) branch gives \(A_f=280.993~\mathrm{MHz}\), a \(0.032\%\) change, and further reduces the average overlap error to \(3.19\times10^{-4}\), with an average final leakage of \(2.89\times10^{-4}\).

The calibrated simulated waveform gives a numerically estimated worst-state overlap error and maximum final leakage of \(1.16\times10^{-3}\). The ideal envelope already gives \(1.08\times10^{-3}\) for both quantities, so the worst-state result is dominated by the selected avoided-crossing trajectory. With local-\(Z\) refitting, the average overlap error remains below \(10^{-3}\) for excursion changes from \(-0.176\%\) to \(+0.154\%\). Conversion of this range to DAC codes requires the device-specific voltage-to-frequency transfer.

At a \(25~\mathrm{ps}\) time step, midpoint and QuTiP propagation produce logical blocks that differ by \(3.8\times10^{-6}\) in relative Frobenius norm after global-phase alignment; increasing each transmon from three to five levels changes the average error by less than \(1.5\times10^{-9}\). This calculation quantifies deterministic envelope-shape sensitivity. Decoherence, flux noise, device-parameter variation, and coupler-specific dynamics require device-specific gate analysis.

Across the tested single-qubit cases, the simulated I/Q path remains below the \(10^{-3}\) worst-state error reference. For the baseband two-qubit envelope, amplitude and local-\(Z\) calibration reduce the average overlap error to \(3.19\times10^{-4}\). Its \(1.16\times10^{-3}\) worst-state value is close to the ideal envelope's \(1.08\times10^{-3}\) and is therefore primarily set by leakage along the selected avoided-crossing trajectory.

\section{Discussion}
\label{sec:discussion}

The results show the benefit and the cost of retaining programmability after optical distribution. The proposed controller shares the high-bandwidth Gaussian and derivative templates within a local group, while amplitude, pulse-enable, and hold-duration states remain channel specific. It is therefore most useful when the same waveform families can be reused across many channels and gate events. Workloads that require independently optimized sample-by-sample waveforms would need additional template rails or complete waveform synthesis. The architectural contribution is the intermediate operating point: more local control than a direct per-channel photonic receiver, without a complete high-speed waveform generator in every 4~K channel.

At \(N=1000\) and \(G=16\), the defined 4~K control-downlink model totals \(0.905~\mathrm{W}\), or approximately 60\% of the representative \(1.5~\mathrm{W}\) stage capacity used in the comparison. All three architectures are compared at the common \(7.4~\mathrm{mV_{pp}}\), \(50~\Omega\) output boundary. The reduction relative to full-waveform Cryo-CMOS comes mainly from replacing three per-channel, \(2~\mathrm{GHz}\) waveform DACs with shared optical templates and an event-rate coefficient DAC. The strict \(50~\Omega\) WDM case lies in the same power range as the proposed controller; impedance matching or a low-power post-detection driver can reduce the WDM receiver power further. The proposed controller accepts this local electrical overhead in return for amplitude and pulse-state updates after distribution. The choice \(G=16\) balances splitter loss against interconnect count and gives 126 template fibers and four shared LO coaxial feeds for the room-temperature-to-4~K downlink. Additional service connections and the 4~K-to-millikelvin wiring enter separately. Spatial fan-out and wavelength packing therefore move the scaling constraint to different parts of the optical network rather than removing it.

The control-chain simulations test whether this reduced waveform-generation model preserves useful gate envelopes. For XY/DRAG control, the two simulated pulse records remain below the \(10^{-3}\) worst-state error reference under one fixed calibration. Separate sweeps place the crossings at approximately \(1.97\%\) common-gain magnitude and \(17.0~\mathrm{dB}\) end-to-end optical-power extinction. The calculation combines two simulated scalar paths as orthogonal quadratures; a physical I/Q channel must also preserve their relative gain, delay, phase, and derivative sign. For the baseband two-qubit path, transferring the ideal-envelope calibration to the simulated S/H waveform gives an average overlap error of \(4.45\times10^{-3}\). Refitting the local-\(Z\) phases reduces it to \(3.62\times10^{-4}\), and recalibrating one frequency-excursion amplitude reduces it further to \(3.19\times10^{-4}\). The remaining \(1.16\times10^{-3}\) worst-state error is close to the \(1.08\times10^{-3}\) ideal-envelope result and is mainly associated with leakage along the selected avoided-crossing trajectory. This closed-system calculation establishes nominal calibratability for the deterministic normalized waveform used in the model. Device-specific performance will additionally depend on measured control-line transfer functions, coherence parameters, and the coupled-device Hamiltonian.

System-level realization depends on synchronization, resonance control, and packaging. The model uses a common room-temperature reference for the optical templates and LO signals, while path-dependent fiber, cable, and on-chip delays require calibration. The \(100~\mathrm{ps}\) timing and \(5.18~\mathrm{mrad}\) phase cases quantify single-channel sensitivity; multi-group deskew and end-to-end latency depend on the implemented clock and command network. After cooldown, the active MRMs and wavelength-selective MRRs require resonance alignment. Nonvolatile phase-change-material trimming has demonstrated a \(0.42~\mathrm{nm}\) cryogenic tuning range without static hold power after programming \cite{adya2025nvRing}, although its write, monitor, and optical-loss overhead must enter the full budget. The defined downlink power model includes signed-\(Q\) switching through the common local-path allowance. A complete 4~K system must additionally account for local digital and timing circuits, photonic tuning, possible LO repeaters, readout, feedback, and package infrastructure. A complete cryogenic local group that measures package loss, timing alignment, the combined I/Q output with signed-\(Q\) restoration, the S/H output, and total power would directly test the predicted trade-off.

\section{Conclusion}
\label{sec:conclusion}

This work shows that low 4~K control power and in-fridge programmability can coexist by using the proposed shared-template photonic/CMOS architecture. Shared optical templates move high-speed waveform synthesis out of 4~K, while local amplitude, pulse-enable, and hold-duration states retain channel-specific control after distribution. At \(N=1000\) and \(G=16\), the defined downlink model gives \(0.905~\mathrm{mW/channel}\), below the normalized full-waveform Cryo-CMOS result and comparable to the strict \(50~\Omega\) WDM case. The simulated XY/DRAG pulses remain below the selected \(10^{-3}\) worst-state error reference. The calibrated S/H example gives \(3.19\times10^{-4}\) average overlap error and \(1.16\times10^{-3}\) worst-state error, placing the representative single- and two-qubit results near the selected feasibility scale. A complete cryogenic local-group implementation will determine the realized power, timing, and control performance.

\ifCLASSOPTIONcaptionsoff
  \newpage
\fi



%




\bibliographystyle{IEEEtran}
\bibliography{references}
\end{document}